\documentclass[%
reprint,
amsmath,amssymb,
aps,
pra,
notitlepage
]{revtex4-1}

\usepackage[english]{babel}
\usepackage[utf8]{inputenc}
\usepackage{amsmath}
\usepackage{physics}
\usepackage{graphicx}
\graphicspath{}
\usepackage[colorinlistoftodos]{todonotes}
\usepackage{lipsum}
\usepackage{multirow}
\usepackage{calc}
\usepackage[bottom]{footmisc}
\usepackage[]{hyperref}
\usepackage{booktabs}
\usepackage[normalem]{ulem}
\usepackage{float}

\addto\captionsenglish{}
\newcommand{\nicefrac}[2]{{#1}/{#2}}

\begin{document}
\title{Interfering distinguishable photons}
\author{Alex~E.~Jones$^{1,2,*}$, Adrian~J.~Menssen$^{1}$, Helen~M.~Chrzanowski$^{1}$, Tom~A.~W.~Wolterink$^{1}$, Valery~S.~Shchesnovich$^{3}$ and Ian~A.~Walmsley$^{1,2}$}
\affiliation{$^{1}$Clarendon Laboratory, Department of Physics, University of Oxford, Oxford, OX1 3PU, United Kingdom,\\
$^{2}$Blackett Laboratory, Imperial College London, London, SW7 2BW, United Kingdom,\\
$^{3}$Center for Natural and Human Sciences, Federal University of ABC, Santo Andr\'e, S\~ao Paulo, 09210-170, Brazil}

\begin{abstract}
One of the central principles of quantum mechanics is that if there are multiple paths that lead to the same event, and there is no way to distinguish between them, interference occurs. It is usually assumed that distinguishing information in the preparation, evolution or measurement of a system is sufficient to destroy interference. For example, determining which slit a particle takes in Young’s double slit experiment or using distinguishable photons in the two-photon Hong-Ou-Mandel effect allow discrimination of the paths leading to detection events, so in both cases interference vanishes. Remarkably for more than three independent quantum particles, distinguishability of the prepared states is not a sufficient condition for multiparticle interference to disappear. Here we experimentally demonstrate this for four photons prepared in pairwise distinguishable states, thus fundamentally challenging intuition of multiparticle interference.
\end{abstract}
\maketitle
\twocolumngrid

The exquisite control of large many-body quantum systems will underpin future quantum technologies. As quantum systems grow in scale, understanding how interference changes with distinguishing information is critical for applications ranging from computing, such as universal quantum computing with photons~\cite{Knill2001} or demonstrating a quantum advantage in boson sampling~\cite{Aaronson2011a}, to simulating particles with exotic statistics~\cite{Matthews2013}. Motivated by intuition from two-photon experiments where distinguishable photons do not exhibit interference, enormous effort is being dedicated to the development of sources of identical single photons of ever increasing quality. While this underlying intuition can be useful, it is an incomplete picture. Here we show that for many-particle systems, state indistinguishability and interference are not synonymous, and counter-intuitively, that photons in separable distinguishable states can interfere. 

How do we quantify the amount of distinguishing information? In the double slit experiment, a single particle may take two paths via the two slits to a point on the detection screen, leading to an interference pattern. As the amount of which-way information on the particle’s trajectory increases, the fringe visibility decreases and reaches zero when an observer could tell with certainty which slit the particle took~\cite{Feynman1963,Wootters1979,Greenberger1988,Mandel1991}. Here the visibility describes the distinguishability of the interfering paths. In quantum erasure experiments, interference is recovered as long as any such information is compensated~\cite{Scully1991,Kim2000,Kwiat1992,Herzog1995}.

For systems of multiple independent particles, additional measures of distinguishability govern the multiparticle interference. The famous Hong-Ou-Mandel (HOM) experiment showed that two identical independent photons incident on the ports of a balanced beam splitter always bunch at the output~\cite{Hong1987}. It is the destructive interference of probability amplitudes for paths where both are transmitted and both are reflected that correlates the photons in this non-classical way. The strength of interference depends on the ability to distinguish these paths, that are here related by an exchange of output port. If the output states associated with each path are indistinguishable under such an exchange, there is complete suppression of coincidences. However if this exchange can in principle provide information on which path has led to a coincidence, the interference weakens and bunching is no longer complete. This is captured by the photons’ pairwise distinguishability – the squared modulus of their quantum state overlap – and is zero (one) for (in)distinguishable states. This motivates the commonly held assumption that no overlap means no interference, and makes the HOM effect ubiquitous as a test of photon indistinguishability~\cite{Pan2012,Brod2019}.

But more particles and more paths allow access to a world of much richer structure owing to the additional exchanges possible~\cite{Greenberger1993,Tillmann2015,Ra2013}. For example, adding a third photon and the possibility of threefold exchange (where all particles permute) introduces a new collective distinguishing parameter -- the triad phase -- arising from the real part of products of state overlaps~\cite{Menssen2017}. Similar phases appear for larger numbers of particles and capture new distinguishing information for higher-order exchanges~\cite{Menssen2017,Shchesnovich2018}.

\begin{figure*}[ht]
\centering
\includegraphics[width=1\textwidth]{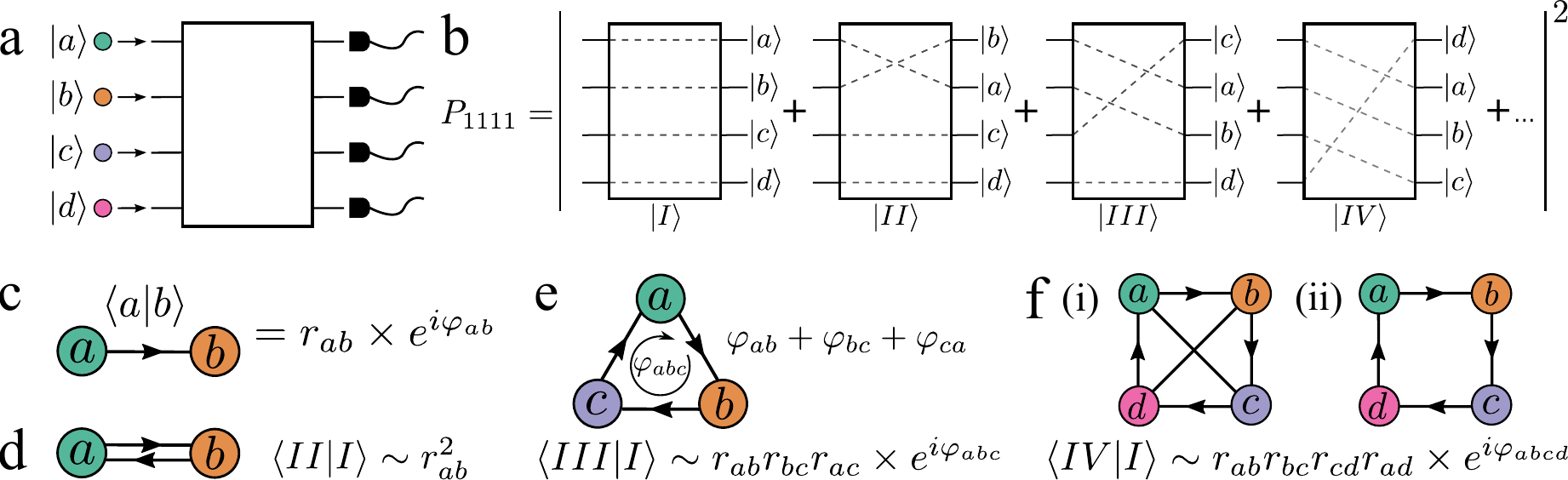}
\caption{\label{fig:fig1}\textbf{Graphs for multiparticle exchanges and distinguishability.} \textbf{a} We inject photons with states labelled $a,b,c,d$ into the inputs of a four-port interferometer and count output coincidences. \textbf{b} The coincidence probability $P_{1111}$ is determined by the probability amplitudes associated with the possible exchange processes that lead to one photon per output. The dependence of these interfering paths on the states' distinguishability can be described by a graph model. \textbf{c} Vertices on a graph represent the states of the photons and directed edges have a weight given by the overlap of states on the connected vertices, here $\braket{a}{b}$. This may be written in modulus-argument form as $r_{ab}\times e^{i\varphi_{ab}}$, and $r_{ab}=0(1)$ for (in)distinguishable states. \textbf{d} This graph has a weight $\braket{a}{b}\times\braket{b}{a}=r_{ab}^2$ and describes the dependence of $P_{1111}$ on the pairwise distinguishability, arising from interference of paths related by pairwise exchange. \textbf{e} The interference of paths related by threefold exchange yields dependence on another distinguishing parameter: the triad phase, given by the sum of the overlaps' arguments $\varphi_{abc}=\varphi_{ab}+\varphi_{bc}+\varphi_{ca}$. We sum the different contributions for each graph: the clockwise and anticlockwise routes, having conjugate phases, will give $\cos\varphi_{abc}$~\cite{Menssen2017}. \textbf{f}~(i) This fourfold exchange has an associated four-particle phase $\varphi_{abcd}$ that can be decomposed into triad phases if all states overlap: here $\varphi_{abcd}=\varphi_{abc}+\varphi_{acd}$. (ii) If pairs of states are distinguishable (\textit{a,c} and \textit{b,d}), the internal edges disappear so $r_{ac}=r_{bd}=0$ and three-particle interference vanishes, but four-particle interference persists~\cite{Shchesnovich2018}.}
\end{figure*}

To investigate the role of distinguishability in multiparticle interference, we consider injecting one photon into each input of a fully-connected interferometer and counting coincidences at the outputs (see Fig.~\ref{fig:fig1}a). The interference due to an $n$-fold exchange contains distinguishing information that may be captured using a graph model~\cite{Shchesnovich2018} (see Fig.~\ref{fig:fig1}b-f). A closed loop drawn on $n$ vertices comprises edges whose weights, when multiplied together, give the corresponding distinguishability dependence of an exchange contribution to interference strength. For $N$ particles whose states all have some mutual overlap, and so a fully connected graph, there will be contributions from loops through $2\leq n\leq N$ vertices. In this situation the pairwise distinguishabilities and triad phases provide a complete specification of the interference pattern~\cite{Menssen2017}.

If, however, pairs of particles are prepared in distinguishable states (Fig.~\ref{fig:fig1}f(ii)), then the triad phase is undefined and three-particle contributions to interference disappear. Yet surprisingly four-particle contributions may persist and multiparticle interference is possible, contrary to intuition from the HOM effect. For example, in general for $N=4$ there may be contributions from two-, three- and four-particle exchanges, giving dependencies on the pairwise distinguishabilities, triad phases and four-particle phases respectively (Fig.~\ref{fig:fig1}). In the case that triad phases are undefined, the lowest order phases determining the interference features are the four-particle ones, given by the sum of the arguments of four state overlaps. For the example shown in Fig. ~\ref{fig:fig1}f(ii), the four-particle phase is $\varphi_{abcd}=\varphi_{ab}+\varphi_{bc}+\varphi_{cd}+\varphi_{da}$. The dependence of multiparticle interference on this phase has not yet been observed. 

\begin{figure}[h!]
	\centering
	\includegraphics[width=0.5\textwidth]{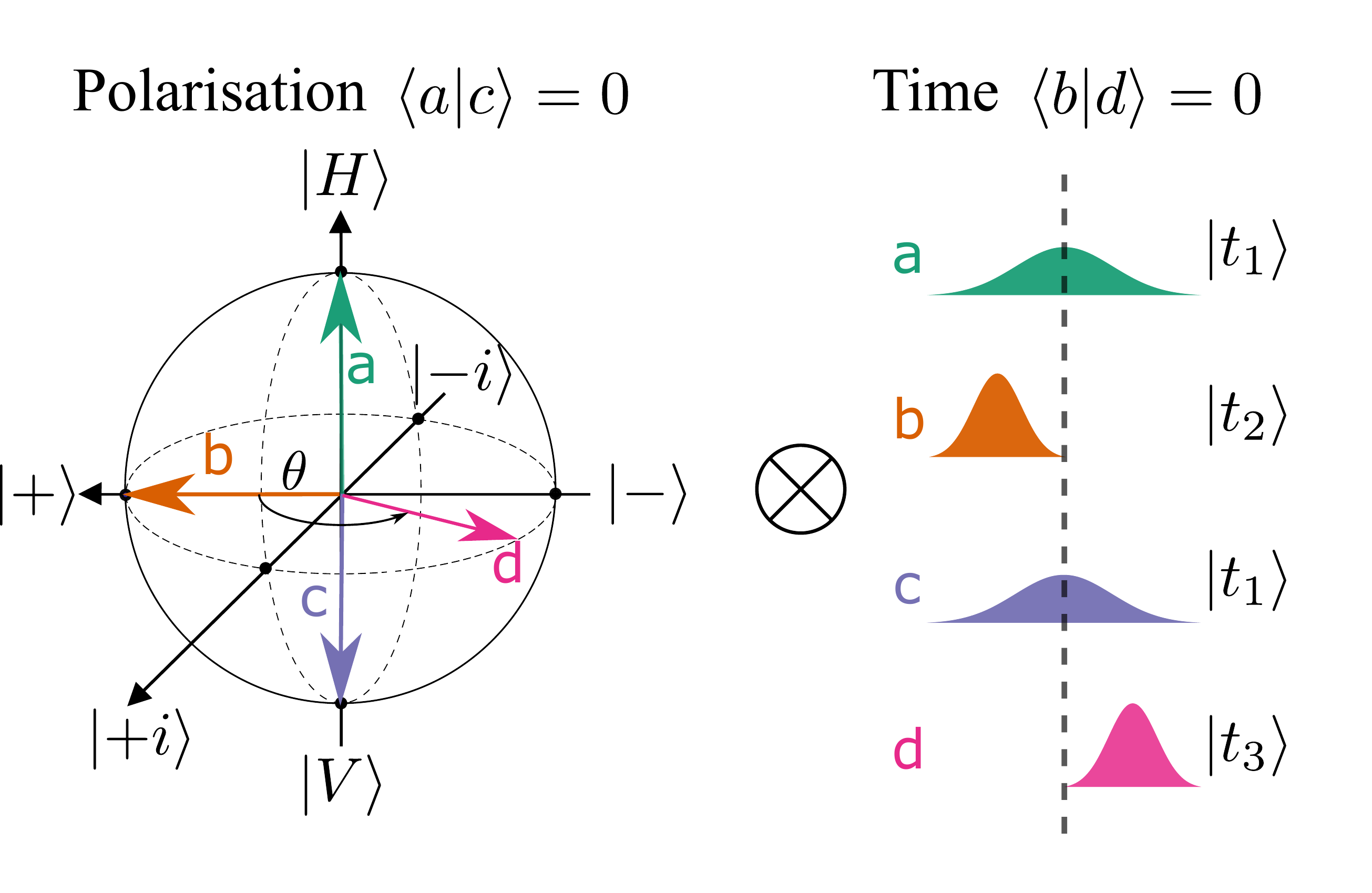}
	\caption{\label{fig:fig2}\textbf{Preparation of four photons in pairwise distinguishable states.} We use polarisation and temporal modes to ensure constant pairwise overlap magnitudes whilst simultaneously eliminating any three-photon overlaps. The four-particle phase is varied by rotating the polarisation of $\ket{d}$ in the equator of the Bloch sphere.}
\end{figure}

In order to, for the first time, experimentally observe multiparticle interference that depends solely on this four-particle phase, we seek a state preparation for four photons that implements the graph in Fig. ~\ref{fig:fig1}f(ii). We eliminate closed paths over three vertices by ensuring that the quantum states are pairwise distinguishable, so $\braket{a}{c}=\braket{b}{d}=0$. We additionally require that all pairwise distinguishabilities $r_{ij}^{2}$ are constant so any variation in coincidences is solely due to the four-particle phase. Varying this phase requires two degrees of freedom: here we use polarisation and temporal modes. We prepare the following states (see Fig.~\ref{fig:fig2}):
\begin{equation}
\begin{aligned}
\label{eqn:statePrep}
\ket{a}&=\ket{H}\otimes\ket{t_{1}},\\
\ket{b}&=\frac{1}{\sqrt{2}}\left(\ket{H}+\ket{V}\right)\otimes\ket{t_{2}},
\end{aligned}
\end{equation}
\begin{equation*}
\begin{aligned}
\ket{c}&=\ket{V}\otimes\ket{t_{1}},\\
\ket{d}&=\frac{1}{\sqrt{2}}\left(\ket{H}+e^{i\theta}\ket{V}\right)\otimes\ket{t_{3}}.
\end{aligned}
\end{equation*}

$\ket{H}$ and $\ket{V}$ denote horizontal and vertical polarisations respectively. $\theta$ may be adjusted using a waveplate to rotate the polarisation of $\ket{d}$ around the equator of the Bloch sphere. The temporal modes $\ket{t_{i}}$ are labelled by the arrival times of the centres of the temporal wavepackets $t_{i}$, and the temporal duration of $\ket{t_{1}}$ is approximately twice that of $\ket{t_{2}}$ and $\ket{t_{3}}$. States $\ket{a}$ and $\ket{c}$ are distinguishable in polarisation, and $\ket{b}$ and $\ket{d}$ are distinguishable in time since $\braket{t_2}{t_3}=0$.

The non-zero state overlaps are:
\begin{equation}
\label{eqn:overlaps}
\begin{aligned}[c]
\braket{a}{b}&=\frac{1}{\sqrt{2}}\braket{t_{1}}{t_{2}},\\
\braket{c}{d}&=\frac{1}{\sqrt{2}}\braket{t_{1}}{t_{3}}e^{i\theta},
\end{aligned}
\hspace{6mm}
\begin{aligned}[c]
\braket{b}{c}&=\frac{1}{\sqrt{2}}\braket{t_{2}}{t_{1}},\\
\braket{d}{a}&=\frac{1}{\sqrt{2}}\braket{t_{3}}{t_{1}},
\end{aligned}
\end{equation}
corresponding to the weights of edges in Fig.~\ref{fig:fig2}f(ii). All these overlaps have constant magnitudes with $\theta$ so two-photon interference terms do not vary. Here only the argument of the polarisation overlap in $\braket{c}{d}$ affects the four-particle phase so $\varphi_{abcd}=\theta$.

\begin{figure*}[!t]
\centering
\includegraphics[width=0.84\textwidth]{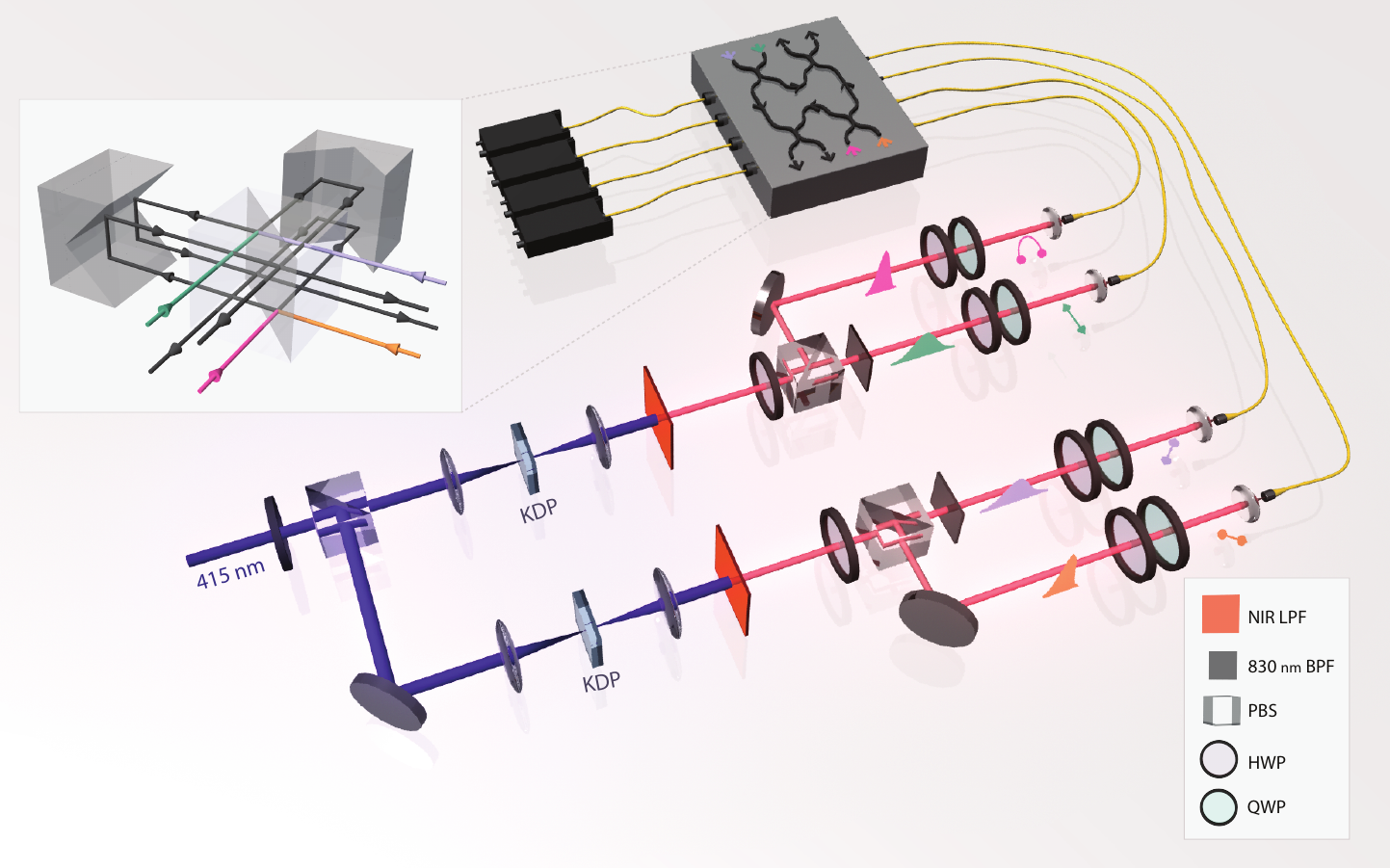}
\caption{\label{fig:fig3}\textbf{Experimental setup.} Two KDP crystals pumped by pulses at 415~nm each probabilistically generate pairs of orthogonally polarised photons by SPDC, and the pump is separated using long-pass filters (LPF). The photons in each pair are spatially separated using polarising beam splitters (PBS), and the signal and idler wavelengths are centred at 830~nm and have bandwidths of 2~nm and 12.5~nm respectively. In order to realise the wavepackets in Fig.~\ref{fig:fig2} we apply additional filtering to the idlers using band-pass filters (BPF) for a bandwidth of around $4.2$~nm. Motorised delay stages apply fixed temporal delays and the polarisation states are prepared using sets of quarter- and half-wave plates (HWP and QWP), and then the photons are coupled into single-mode fibres to the interferometer. We use a free-space bulk-optic four-mode interferometer: retroreflecting mirrors are used to double pass spatial modes through a single balanced dielectric beam splitter, allowing a folded design with straightforward path-length matching and good stability (see inset and Supplementary Material). Single photon avalanche photodiodes detect photons at the four quitter outputs and coincidences are recorded using a commercial time tagger with 4~ns coincidence window.}
\end{figure*}

The experimental setup is shown in Fig.~\ref{fig:fig3}. We generate four photons using a pair of identical spontaneous parametric down-conversion (SPDC) sources based on bulk potassium dihydrogen phosphate (KDP) crystals. These are pumped by a frequency-doubled Ti:Sapphire femtosecond laser pulsing at 80 MHz and birefringent phase-matching allows generation of spectrally factorable pairs of signal and idler photons in degenerate type-II collinear SPDC~\cite{Mosley2008}. We prepare the photons' states as in equation (\ref{eqn:statePrep}) (Fig.~\ref{fig:fig2}) and interfere them in a balanced four-mode interferometer (`quitter', see inset of Fig.~\ref{fig:fig3}) described by the unitary matrix
\begin{equation}
\label{eqn:quitterUnitary}
U_{quit}=\frac{1}{2}
\begin{pmatrix}
1&1&1&1\\
1&1&-1&-1\\
1&-1&e^{i\chi}&-e^{i\chi}\\
1&-1&-e^{i\chi}&e^{i\chi}
\end{pmatrix}.
\end{equation}
The elements of this matrix are the couplings between the input and output mode operators, where the rows correspond to inputs 1-4 and columns to outputs 5-8. $\chi$ is a phase that does not affect the splitting ratio. The pairwise distinguishable states of equation~\ref{eqn:statePrep}, labelled \textit{a, b, c} and \textit{d}, are injected into ports 4, 2, 3 and 1 respectively, and the coincidence probability at the quitter's outputs is
\begin{widetext}
\begin{equation}
\label{eqn:P1111}
P_{1111}=\frac{1}{32}\big(3-r_{ab}^{2}-r_{bc}^{2}-r_{cd}^{2}-r_{ad}^{2} +(\cos 2\chi+2)\times(r_{ab}^{2}r_{cd}^{2}+r_{ad}^{2}r_{bc}^{2})+2(\cos 2\chi-2)\times r_{ab}r_{bc}r_{cd}r_{ad}\cos\varphi_{abcd}\big).
\end{equation}
\end{widetext}

The terms in the pairwise distinguishabilities $r_{ij}^{2}$ are HOM-type contributions. The absence of $r_{ac}$ and $r_{bd}$ is due to the enforced pairwise distinguishability and also means no triad phases appear. Terms given by products of two pairwise distinguishabilities correspond to coverings of the graph in Fig.~\ref{fig:fig1}f(ii) using two closed loops over two pairs of vertices (each loop similar to that of Fig.~\ref{fig:fig1}d). The last term is the four-photon exchange with a dependence on $\varphi_{abcd}$ and is the only exchange contribution that varies for our state preparation. Its relative strength depends on the phase $\chi$ that varies during the experiment but is measured independently (see Supplementary Material). For ideal states these moduli $r_{ij}=1/2$ and then, assuming $\chi=\pi/2$ for maximum variations with $\varphi_{abcd}$, the signal would have a visibility of $30\%$ (defined as (max-min)/max).

\begin{figure*}[t]
	\centering
	\includegraphics[width=0.835\textwidth]{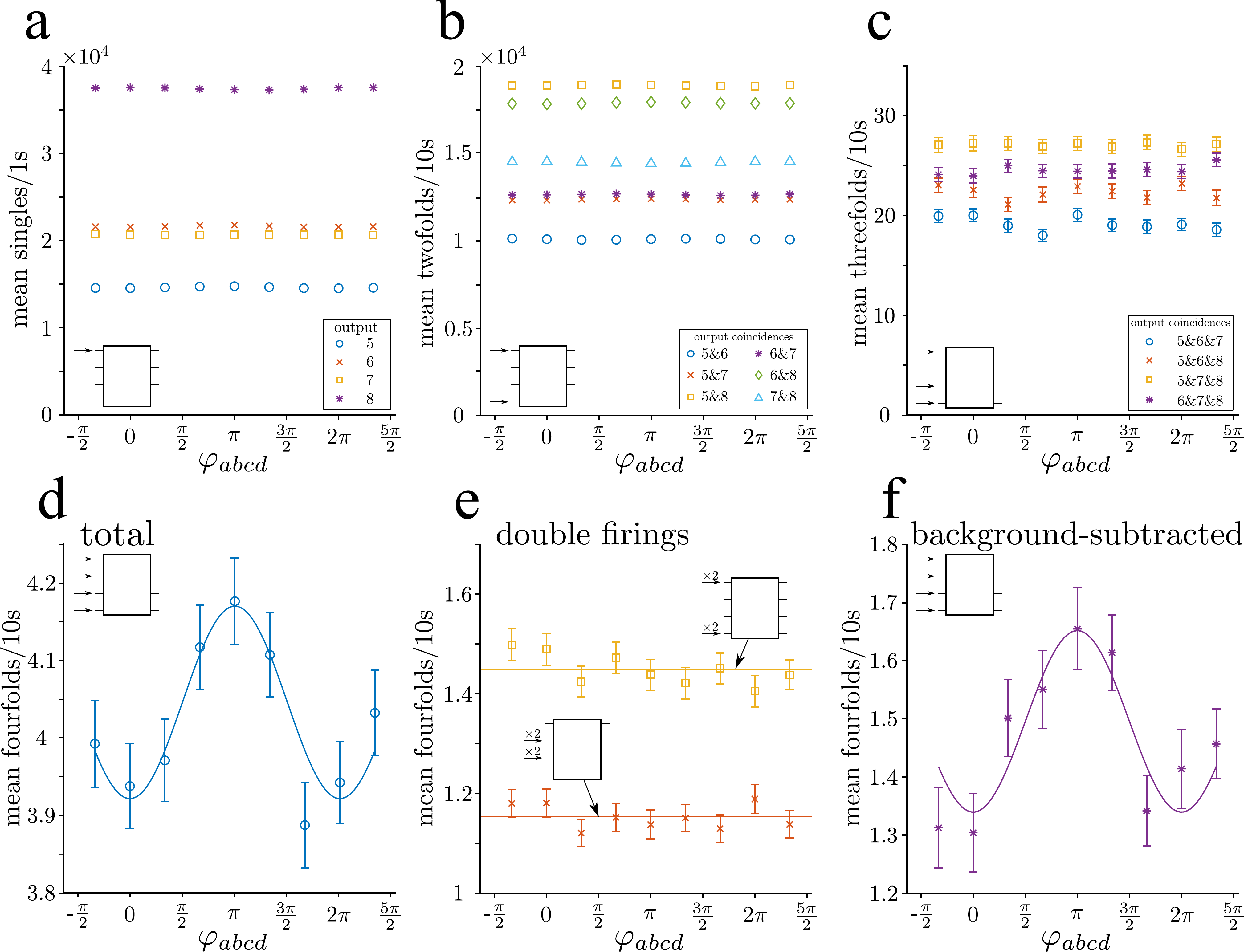}
	\caption{\label{fig:fig4}\textbf{Counts measured at the output ports of the quitter interferometer.} \textbf{a} Output singles when sending $\ket{d}$ into the first quitter input and blocking all other inputs. \textbf{b} Twofold output coincidences when sending states $\ket{d}$ and $\ket{a}$ into the first and fourth quitter inputs. \textbf{c} Threefold output coincidences when sending states $\ket{d},\ket{c}$ and $\ket{a}$ into the first, third and fourth quitter inputs. \textbf{d} Total fourfold output coincidences when the emissions from both sources are sent to the quitter. The fitted cosine has a visibility of $5.9\pm1.0\%$ and the numbers of counts per data point are between 6,455 and 6,012. \textbf{e} By closing one of the SPDC sources, we record the output fourfolds arising from double emissions by the other source. These signals have an average of 2,160 and 1,740 counts per point. \textbf{f} Subtracting the backgrounds in \textbf{e} from the fourfold counts in \textbf{d} yields the $P_{1111}$ signal showing the expected $-\cos\varphi_{abcd}$ behaviour of equation~\ref{eqn:P1111}. The fitted cosine has a visibility of $18.9\pm3.7\%$ and counts per point of between 2,717 and 2,216. Error bars from repeated sweeps are omitted if smaller than the data markers. See Supplementary Material for details of the data analysis, background subtraction and simulations, and the main text for discussion of these plots.}
\end{figure*}

When the emission from both sources is injected into the quitter, it is impossible to know whether a detection of a fourfold output coincidence is due to each source firing once or one of the sources firing twice. The former leads to the desired $P_{1111}$ signal whilst the latter lead to fourfolds depending only on $r_{ad}$ or $r_{bc}$, depending on which source fires twice. These double firings therefore contribute a flat background to the total fourfold signal that can be measured separately and subtracted from the total signal.

The four-particle phase $\varphi_{abcd}$ is varied by rotating the polarisation of $\ket{d}$. At each setting, all singles and two-, three-, and fourfold coincidence counts are recorded at the quitter outputs when opening all different combinations of inputs. This allows independent measurement of variations caused by lower-order interference contributions and of the background from source double emissions, and results are shown in Fig.~\ref{fig:fig4}.

Fig.~\ref{fig:fig4}a shows the mean singles measured at the quitter outputs when injecting $\ket{d}$ into the quitter and blocking all other inputs. The polarisation of this state is the only parameter that is changed during the experiment. These counts exhibit an average variation of below $0.5\%$ that arises from slight polarisation-dependence of the quitter. Singles counts were also recorded when opening the other source arms individually and are all constant with $\varphi_{abcd}$.

In Fig.~\ref{fig:fig4}b,c we plot the output two- and threefold coincidences recorded for a subset of input configurations involving $\ket{d}$. The former exhibit an average of $0.2\%$ variation and the latter are constant within error. Together with additional measurements for the other input combinations presented in the Supplementary Material, we confirm that any contributions to $P_{1111}$ from pairwise distinguishabilities or triad phases are far smaller than the $\varphi_{abcd}$-dependent four-photon exchange term.

In Fig.~\ref{fig:fig4}d we plot the output fourfold coincidences recorded when both SPDC sources are injected into the quitter. This signal contains the desired $P_{1111}$ term -- corresponding to each source firing once -- and also contributions from each source firing twice. The result is a cosine variation with a visibility of $5.9\pm1.0\%$, compared to a predicted $6.7\%$ that accounts for higher-order emissions of up to six photons, input losses, residual spectral distinguishability and the sampled $\chi$ values. The double emissions are shown in Fig.~\ref{fig:fig4}e and are constant with $\varphi_{abcd}$, as expected since they depend only on pairwise distinguishabilities. Subtracting these constant backgrounds from the total fourfolds yields the signal in Fig.~\ref{fig:fig4}f. This corresponds to our measurement of $P_{1111}$ and has a fitted visibility of $18.9\pm3.7\%$, consistent with a predicted value of $17.9\%$. Reduction from the ideal $30\%$ is mainly due to variations of the quitter phase $\chi$ and imperfect state preparation. We also postselect for data recorded when $\chi\approx\pi/2$ and find an increased visibility of $23.6\pm7.0\%$, consistent with a predicted value of $22.5\%$. The variations in total and background-subtracted fourfolds observed in Fig.~\ref{fig:fig4}d,f cannot be attributed to variations in the lower-order exchange contributions. The cosine variation arises from a dependence on the four-particle phase $\varphi_{abcd}$ defined by our preparation of pairwise distinguishable states.

This is the first evidence of multiparticle interference of four photons prepared in separable pairwise distinguishable states. It shows that distinguishability of quantum states is not always accompanied by a loss of interference: multiparticle interference can persist.

The effect demonstrated here extends to more particles (by analogy with Fig.~\ref{fig:fig1}) but with a smaller signal visibility due to decreasing state overlaps~\cite{Shchesnovich2018}. The ability to prepare a state of the electromagnetic field that exhibits $N$-fold photon correlations without any in the lower orders is not unique to the separable states we have presented here. Approaches using entangled states of light have been proposed~\cite{Rice1994}, and shown experimentally for three photons~\cite{Agne2017}.

Interference of independent distinguishable photons is also possible with careful control of the measurement process: photons with orthogonal polarisations can interfere if polarisers are placed in front of the detectors, and spectrally distinguishable photons can interfere if the detectors have high timing resolution~\cite{Legero2003,Laibacher2015,Wang2018b}. However our experiment uses an input state comprising independent, pairwise distinguishable photons and detectors that resolve only the presence of a photon, not the internal mode structure. It is the indistinguishability of a subset of paths corresponding to possible exchange processes in an interferometer that means interference is possible despite distinguishable states.

As quantum systems continue to be scaled up, it is important to remember the subtleties of interference and the need for careful consideration when generalising observations at smaller scales. Both the double slit experiment and the HOM effect compelled physicists to revise what qualifies as intuition in the quantum world: distinguishability appeared to be accompanied by a loss of interference and a return to classical behaviour. We have shown that, on introducing another pair of independent photons to the famous HOM effect, interference effects are possible that contradict even this long-held intuition.

\section*{Acknowledgements}
The authors thank Thomas Hiemstra for help designing the interferometer. This work was supported by the UK Engineering and Physical Sciences Research Council (EPSRC EP/K034480/1). AEJ is supported by the EPSRC via the Controlled Quantum Dynamics CDT (EP/L016524/1). AJM is supported by the James Buckee scholarship from Merton College. VSS is supported by CNPq  and FAPESP of Brazil. TAWW is supported by the Fondation Wiener-Anspach. IAW acknowledges an ERC Advanced Grant (MOQUACINO). 
\section*{Author contributions}
AEJ performed the experiment, modelling and data analysis with assistance from AJM and TAWW. AJM, VSS and AEJ developed the theory and conceived the experiment. HMC and IAW supervised the project. AEJ wrote the manuscript with input from all authors.

\bibliographystyle{hphysrev}
\renewcommand{\bibpreamble}{*a.jones14@imperial.ac.uk}


\newpage
\newpage
\onecolumngrid
\newpage

\section*{Supplementary material}
\section{Transition probabilities in a scattering experiment}
\label{app:transitionProbs}
The probability of observing a counting pattern $\vec{s}$ given an input configuration $\vec{r}$, associated pure state vectors $\ket{\phi_{i}}$, and scattering unitary $U$ is given by~\cite{Tichy2015}:
\begin{equation}
\label{eqn:generalProb}
P_{\mathrm{pure}}(\vec{r},\vec{s})=\mathcal{N}\sum_{\rho\in S_n}\bigg[\prod_{j=1}^{n}\mathcal{S}_{j,\rho_j}\bigg]\times\mathrm{perm}\big(M*M^{*}_{\rho}\big).
\end{equation}
$\mathcal{N}=1/\prod_{j}r_{j}!s_{j}!$ is a normalisation factor, $\mathcal{S}_{ij}=\braket{\phi_i}{\phi_j}$ is the Hermitian distinguishability matrix whose elements are the state overlaps, and $M$ is the effective scattering matrix obtained by selecting rows and columns from the unitary $U$ with multiplicities given by the occupation numbers of $\vec{r}$ and $\vec{s}$ respectively. $M^*_{\rho}$ is the element-wise complex conjugate of $M$ with its rows permuted according to the corresponding element $\rho$ of the symmetric group $S_n$. The product $M*M_{\rho}^*$ is meant element-wise, and `perm' is the matrix permanent.

\section{Four-mode interferometer}
\label{app:quitter}
We use a fully connected balanced four mode splitter called a ``quitter" that is described by the unitary matrix:
\begin{equation}
\label{eqn:appQuitterUnitary}
U_{quit}=\frac{1}{2}\left(\begin{matrix}
1&1&1&1\\
1&1&-1&-1\\
1&-1&e^{i\chi}&-e^{i\chi}\\
1&-1&-e^{i\chi}&e^{i\chi}
\end{matrix}\right),
\end{equation}
with rows corresponding to inputs 1-4 and columns to outputs 5-8. All matrix element magnitudes are equal and so all input spatial modes are equally coupled. There is a free internal phase $\chi$ that affects the four-photon exchange contribution (see Supplementary~\ref{app:measuringQuitterPhase}). The interferometer's operation should be independent of polarisation and this is discussed in Supplementary~\ref{appxSec:dataNorm}.

\begin{figure}[h]
	\centering
	\includegraphics[width=0.95\textwidth]{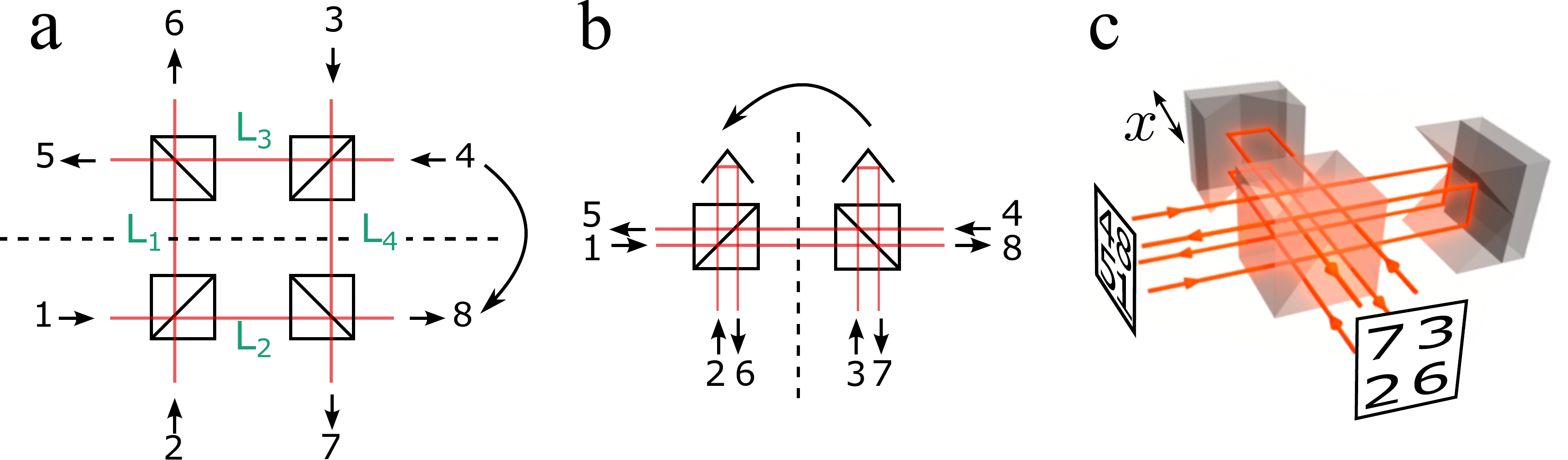}
	\caption{\label{fig:appFoldedQuitter}\textbf{a} Bulk optic realisation of a balanced four-port interferometer. All beam splitters have 50\% reflectivity, and the numbers 1-4 label input ports and 5-8 label the outputs. Internal path lengths are labelled $L_i$. We fold the interferometer along the dashed line by using retroreflectors to double-pass through a pair of beam splitters at different lateral positions to give \textbf{b}. We then fold again using retroreflectors that translate the beams to a different vertical position out of the page as shown. This results in the configuration in \textbf{c} where the beams pass through a single 50:50 beam splitter multiple times, simplifying alignment and improving stability. Lengths $L_{1}$ and $L_{4}$ can be changed using a delay stage on the lateral retroreflector (labelled $x$). The corresponding input and output ports are labelled on the free faces of the beam splitter.}
\end{figure}

Such a device has been investigated theoretically for testing the distinguishability of four particles~\cite{Tichy2011,Shchesnovich2018}. Bulk optic realisations have previously been used for Bell state analysis and two-photon interference experiments~\cite{Michler1996,Mattle1995}. Here we begin with such constructions using four balanced beam splitters, and then simplify alignment and path length matching by folding the interferometer twice using retroreflectors (Fig.~\ref{fig:appFoldedQuitter}a-c).

The internal path lengths $L_i$ must be matched to within the coherence length of the photons: $L_1\approx L_3$ and $L_2\approx L_4$. This is achieved by mounting the lateral retroreflector on a delay stage and shifting its position $x$ as shown in Fig.~\ref{fig:appFoldedQuitter}c, allowing $L_1\rightarrow L_1+2x$ and $L_4\rightarrow L_4+2x$. The internal phase $\chi$ is determined by path length mismatch at the scale of the central wavelength $\lambda_{0}$ by:
\begin{equation}
\label{eqn:quitterPhase}
\chi=\frac{2\pi}{\lambda_{0}}\times\big((L_1+L_4)-(L_2+L_3)\big).
\end{equation} 

\section{Measuring the quitter phase}
\label{app:measuringQuitterPhase}
\begin{figure}[h]
	\centering
	\includegraphics[width=0.5\textwidth]{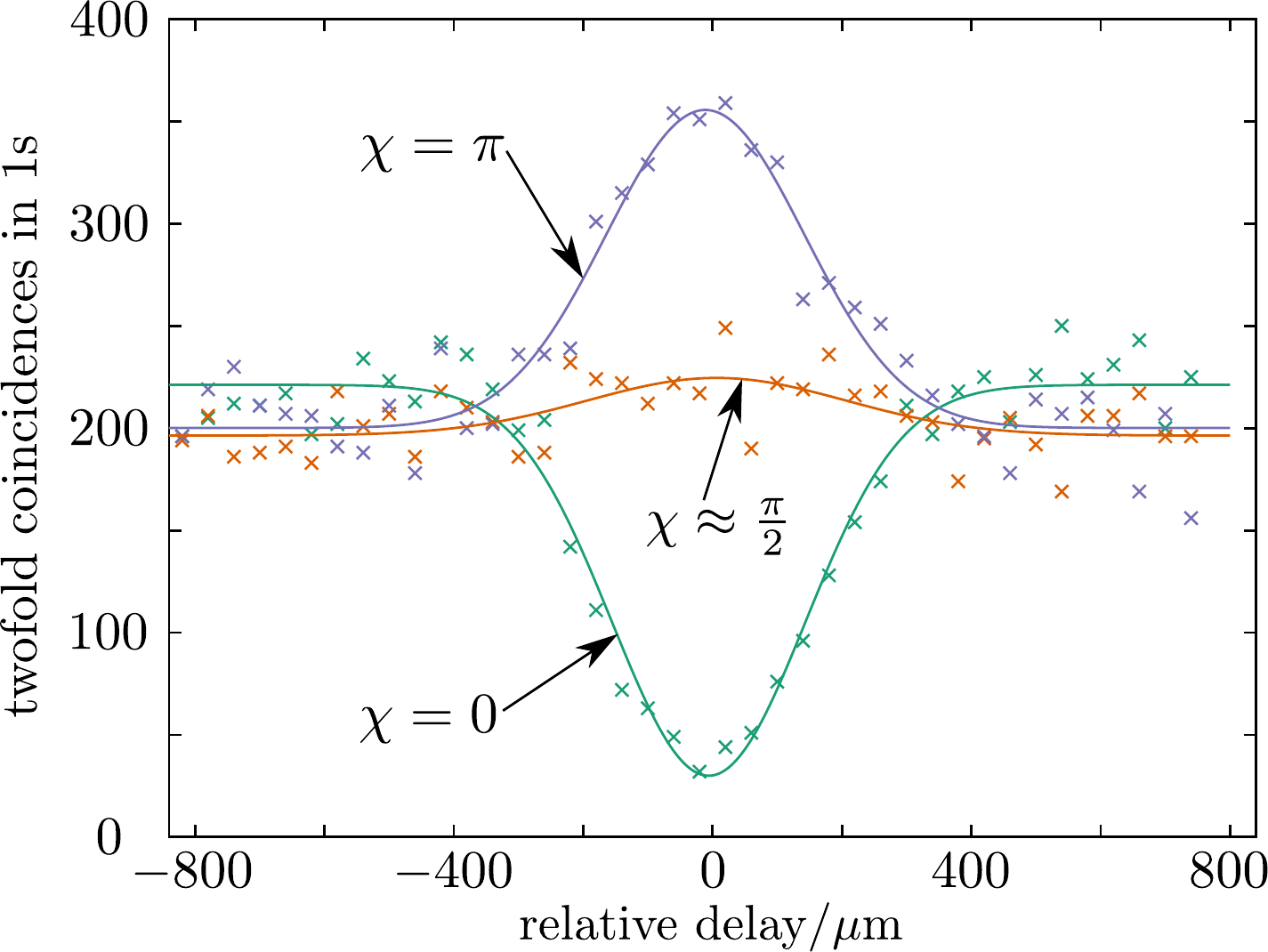}
	\caption{\label{fig:phaseDepHOM}Hong-Ou-Mandel dips through the bulk quitter, demonstrating the dependence on the internal phase $\chi$.}
\end{figure}

There is no single-particle interference in the device so we use two-photon interference to monitor the interferometer phase $\chi$. If two photons with state vectors $\ket{\alpha},\ket{\beta}$ are injected into inputs 2 and 3, and then outputs 5 and 7 are monitored for coincidences, the probability is (using equation~\ref{eqn:generalProb}):
\begin{equation}
\label{eqn:phaseDepHOM}
P^{2,3}_{5,7}=\frac{1}{8}(1-r_{\alpha\beta}^2\times\cos\chi),
\end{equation}
where $r_{\alpha\beta}=\lvert\braket{\alpha}{\beta}\rvert$. From Fig.~\ref{fig:appFoldedQuitter}a we see that this phase-dependence arises despite the two photons never meeting, highlighting that it is the interference of paths to an event that determines statistics~\cite{Laing2012,Branning1999}. We measured Hong-Ou-Mandel dips using a pair of photons with $r_{\alpha\beta}^2\approx0.9$ to verify this phase dependence (Fig.~\ref{fig:phaseDepHOM}).

Given ideal pairwise distinguishable states labelled $\textit{a,b,c,d}$ (where $\braket{a}{c}=0$ and $\braket{b}{d}=0$), the visibility of the fourfold coincidence probability at the quitter outputs is maximised for the following injection order:
\begin{equation}
\begin{aligned}
\label{eqn:stateOrder}
\ket{a}&\rightarrow \mathrm{input}\enskip4,\\
\ket{b}&\rightarrow \mathrm{input}\enskip2,\\
\ket{c}&\rightarrow \mathrm{input}\enskip3,\\
\ket{d}&\rightarrow \mathrm{input}\enskip1.
\end{aligned}
\end{equation}
With this configuration the fourfold coincidence probability (equation~\ref{eqn:P1111} in main text, found using the expression for scattering probabilities in equation~\ref{eqn:generalProb}) is:
\begin{equation}
\label{eqn:appP1111}
P_{1111}=\frac{1}{32}\left(3-r_{ab}^{2}-r_{bc}^{2}-r_{cd}^{2}-r_{ad}^{2} +(\cos 2\chi+2)\times(r_{ab}^{2}r_{cd}^{2}+r_{ad}^{2}r_{bc}^{2})+2(\cos 2\chi-2)\times r_{ab}r_{bc}r_{cd}r_{ad}\cos\varphi_{abcd}\right).
\end{equation}

\begin{figure}[h]
	\centering
	\includegraphics[width=0.5\textwidth]{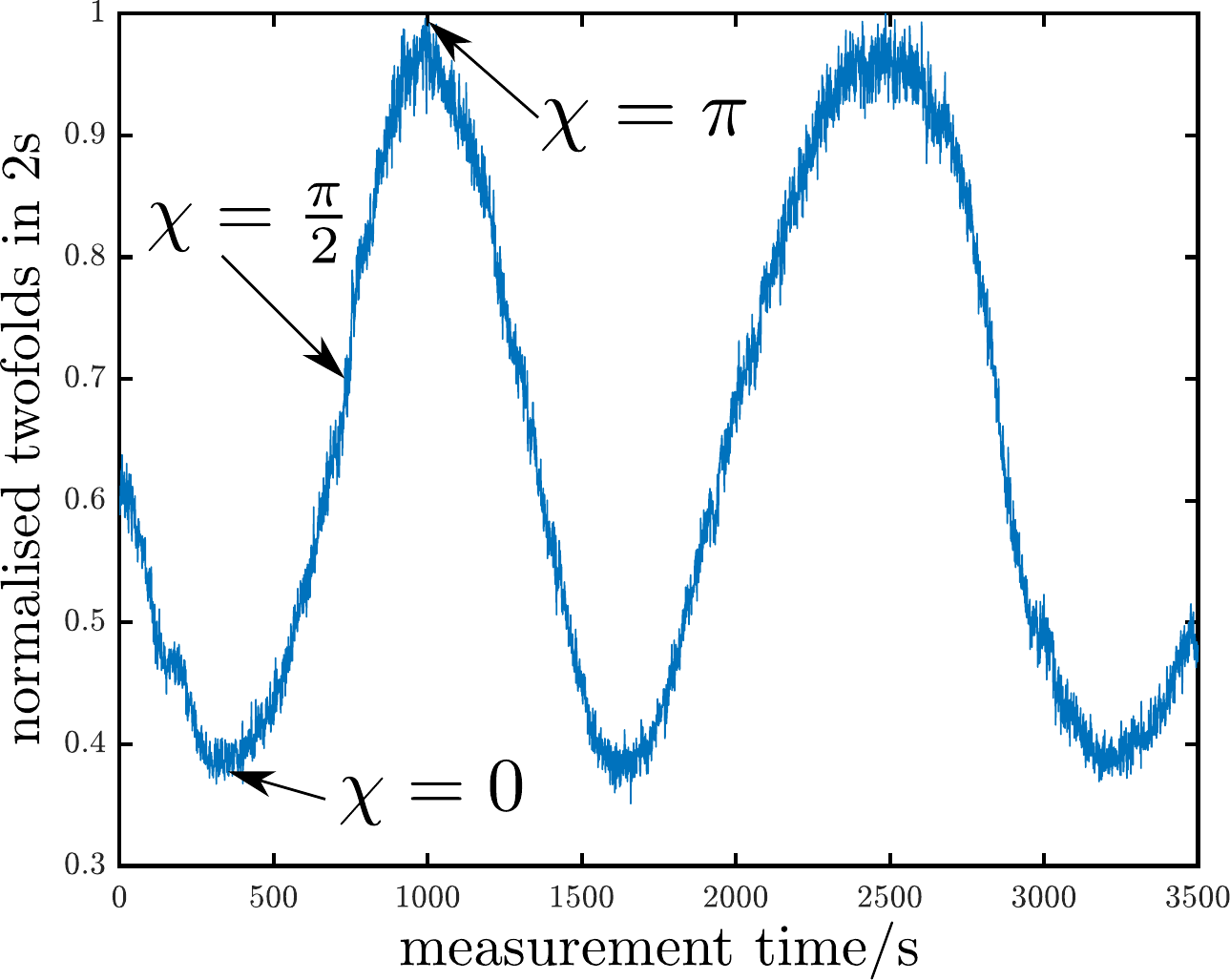}
	\caption{\label{fig:lockingDemo}$\chi$-dependent $P^{2,3}_{5,7}$ two-photon coincidences at a fixed temporal delay, scaled to eliminate the effect of variations in singles counts and then normalised to the maximum value. The slow variations are due to the refractive index of air changing with humidity and temperature variations. This signal can be used to determine which coincidences correspond to $\chi=0$ and $\chi=\pi$, then equation~\ref{eqn:phaseDepHOM} allows inference of $\chi$ throughout the experiment.}
\end{figure}

Since $P_{1111}$ depends on $\chi$, it is important to measure it during the experiment. Before recording the desired fourfold coincidences at each value of $\varphi_{abcd}$ when both sources are open, one source is blocked and the other is used to measure twofold coincidences at a fixed temporal and polarisation overlap (in the experiment we use constant $r_{bc}^{2}$). The phase $\chi$ may then be inferred using equation~\ref{eqn:phaseDepHOM} and, without any adjustments, it drifts with temperature and humidity in the laboratory (see Fig.~\ref{fig:lockingDemo}), from $\chi=\pi$ giving maximum twofolds to $\chi=0$ giving minimum twofolds.

The $P_{1111}$ signal visibility with $\varphi_{abcd}$ is maximised when $\chi=(2n+1)\pi/2$ and minimised when $\chi=n\pi$, giving $30\%$ and $10\%$ respectively for ideally pairwise distinguishable states labelled \textit{a,b,c,d}, where all non-zero overlap magnitudes are $1/2$ (see Supplementary~\ref{app:distStatePrep} for discussion of experimentally realised states). Uniformly sampling $\chi$ would therefore lead to $20\%$ visibility but we use a basic locking procedure to increase the time spent recording fourfolds at $\chi\approx\pi/2$: the vertical retroreflector of the folded quitter is attached to a piezo-controlled mount and a small tilt permits some variation of $(L_2+L_3)\rightarrow(L_2+L_3)\pm\epsilon$. A single piezo step corresponds to $\epsilon\approx\nicefrac{\lambda}{20}$. By monitoring the twofold coincidences as in Fig.~\ref{fig:lockingDemo}, the path length can be actively changed to spend more time near $\chi=\pi/2$ and so increase the signal visibility. If the rate of change of $\chi$ due to ambient drifts is comparable to the frequency of applying the procedure, there is a risk of beam steering that could affect fibre coupling. Therefore this technique was only used when ambient drifts were slow: the result is that around half the recorded data are obtained for $\arccos\left(\frac{1}{3}\right)=1.23\leq\chi\leq\pi-\arccos\left(\frac{1}{3}\right)=1.91$, where $\chi$ is constrained to vary across one third of the total range defined by $\chi=0,\pi$ in Fig.~\ref{fig:lockingDemo}, and the other half is measured when $\chi$ is in the remaining range, where the $P_{1111}$ visibility is lower. A list of sampled $\chi$ values is recorded during the experiment and used in a simulation described in Supplementary~\ref{app:subtractingStatistics} to predict the fourfold coincidence signals' visibilities.

\section{Further details of distinguishable state preparation}
\label{app:distStatePrep}
\begin{figure}[h]
	\centering
	\includegraphics[width=0.475\textwidth]{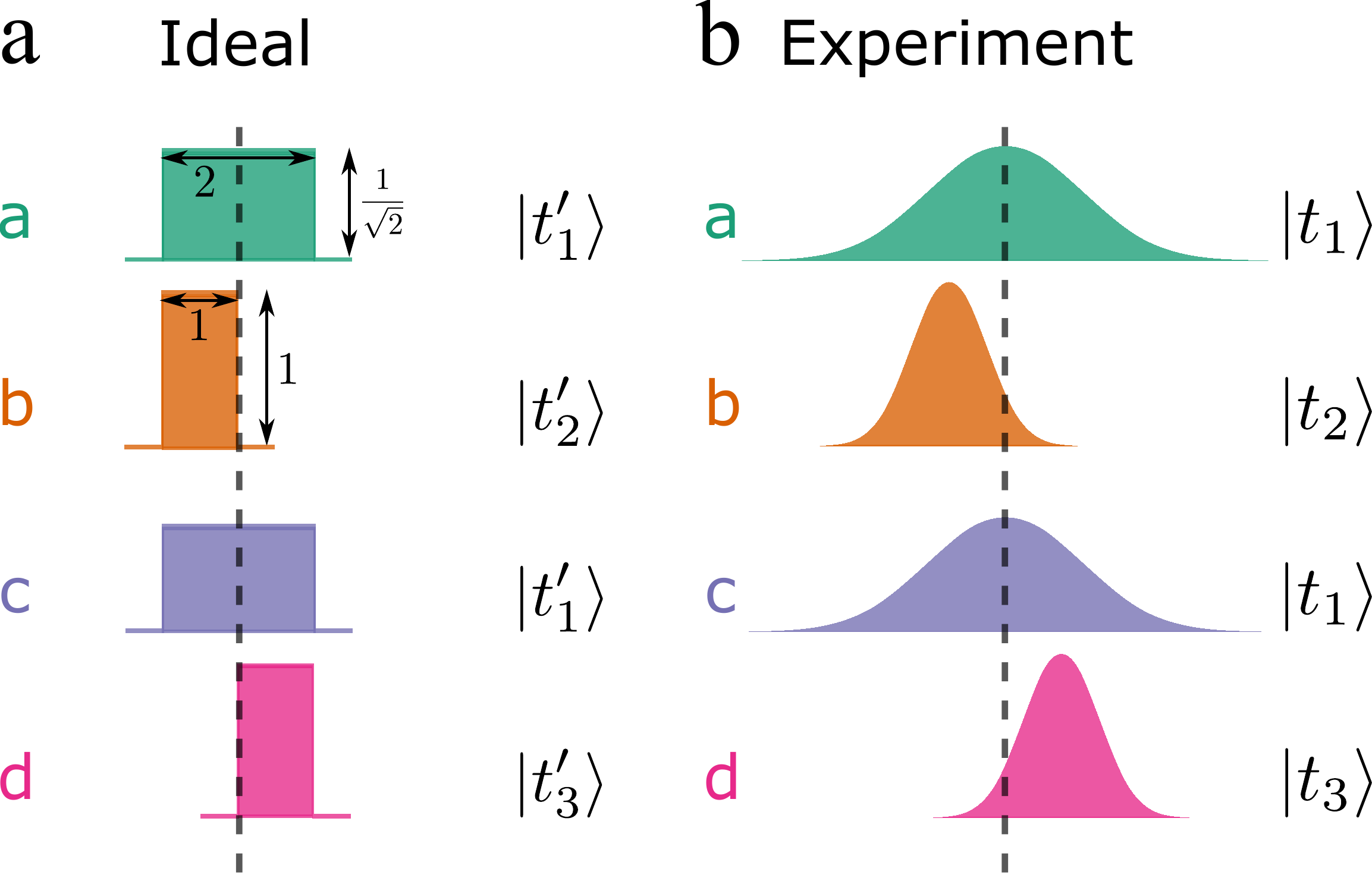}
	\caption{\label{fig:temporalModes}Ideal and experimentally realised temporal mode configurations. \textbf{a} Top-hat functions with the indicated relative durations and amplitudes would be ideal to maximise $\braket{t_{1}'}{t_{2}'},\braket{t_{1}'}{t_{3}'}$, giving a large prefactor for $\cos\varphi_{abcd}$ variations whilst also enforcing $\braket{t_{2}'}{t_{3}'}=0$. \textbf{b} In practice our parametric down-conversion sources generate photons with a near Gaussian spectral, and therefore temporal, profile. The filtered temporal modes $\{\ket{t_{i}}\}$ in the experiment have a $\sim$2.1:1 duration ratio -- shown to scale in this picture -- and do not achieve perfect distinguishability (see Supplementary~\ref{app:additionalExchangeContributions} for discussion).}
\end{figure}

In order to isolate the effects of distinguishable state interference, we prepare four photons with the internal states (equation~\ref{eqn:statePrep} in the main text):
\begin{equation}
\begin{aligned}
\label{eqn:appStatePrep}
\ket{a}&=\ket{H}\otimes\ket{t_{1}},\\
\ket{b}&=\ket{+}\otimes\ket{t_{2}},\\
\ket{c}&=\ket{V}\otimes\ket{t_{1}},\\
\ket{d}&=\ket{\theta}\otimes\ket{t_{3}}.
\end{aligned}
\end{equation}
The polarisation state $\ket{\theta}=\frac{1}{\sqrt{2}}(\ket{H}+e^{i\theta}\ket{V})$ rotates in the equator of the Bloch sphere. States $\ket{a}$ and $\ket{c}$ are orthogonal in polarisation so $\braket{a}{c}=0$. Now we would like $\ket{b}$ and $\ket{d}$ to be distinguishable in temporal mode so want $\braket{t_2}{t_3}=0$. This could ideally be achieved by, for example, using wavepackets $\{\ket{t'_{i}}\}$ with top-hat profiles as in Fig.~\ref{fig:temporalModes}a (assuming flat spectral phase). One could then write out the ideal non-zero scalar products as:
\begin{equation}
\begin{aligned}
\label{eqn:appOverlaps}
\braket{a}{b}&=\braket{H}{+}\times\braket{t'_{1}}{t'_{2}}=\frac{1}{2},\\
\braket{b}{c}&=\braket{+}{V}\times\braket{t'_{2}}{t'_{1}}=\frac{1}{2},\\
\braket{c}{d}&=\braket{V}{\theta}\times\braket{t'_{1}}{t'_{3}}=\frac{1}{2}e^{i\theta},\\
\braket{d}{a}&=\braket{\theta}{H}\times\braket{t'_{3}}{t'_{1}}=\frac{1}{2}.
\end{aligned}
\end{equation}
Substituting these into the equation for the four-photon coincidence probability of equation~\ref{eqn:P1111} in the main text gives ideal visibilities of $P_{1111}$ with $\varphi_{abcd}$ of between $30\%$ and $10\%$ for $\chi=(2n+1)\pi/2$ and $\chi=n\pi$ respectively. Triad phase contributions are eliminated and two-particle interferences are constant.

In practice our SPDC sources generate photons in transform-limited near Gaussian wavepackets that, without spectral filtering, have temporal durations in the ratio $\sim$5:1. Their tails mean that perfect distinguishability is difficult to achieve whilst also maintaining appreciable overlap magnitudes $\abs{\braket{t_{1}}{t_{2}}},\abs{\braket{t_{1}}{t_{3}}}$: these affect the moduli $r_{ab}$ and $r_{ad}$ that determine the size of the four-particle contribution in $P_{1111}$. We can adjust the relative durations of the wavepackets by filtering the spectrum and can also adjust the relative arrival times $t_{2}$ and $t_{3}$. It is then a case of balancing the filtered count rates, the visibility of the fourfold coincidence probability, and the effect of undesired triad phase terms appearing because $\braket{b}{d}\neq 0$. 

A compromise leads us to say that the two temporal modes are ``distinguishable'' if the Hong-Ou-Mandel dip visibility on a balanced beam splitter is at most 1\%, or equivalently $\abs{\braket{t_{2}}{t_{3}}}\leq0.1$. We apply spectral filtering to achieve a temporal duration ratio of $\sim$2.1:1 (see spectra in Fig.~\ref{fig:fourPhotSpectra}) and then the temporal modes $\ket{t_{2}}$ and $\ket{t_{3}}$ are walked off each other using this distinguishability criterion to achieve the configuration $\{\ket{t_{i}}\}$ shown in Fig.~\ref{fig:temporalModes}b (see Supplementary~\ref{app:exptPreparation} for experimental details).

The Gaussian wavepackets delayed by time $t_{i}$, with central frequency $\Omega$ and variance in time $\sigma_{i}^{2}$ are:
\begin{equation}
\ket{t_{i}}=\left(\pi\sigma_i^2\right)^{-\frac{1}{4}}\int \mathrm{d}\tau \mathrm{exp}\bigg(-\frac{(t_{i}-\tau)^2}{2\sigma_i^2}+i\Omega(t_{i}-\tau)\bigg)\ket{\tau}.
\end{equation}
The products of cycles of temporal overlaps for transform-limited Gaussian temporal wavepackets are real~\cite{Menssen2017}. The argument of $\braket{c}{d}$ is therefore controlled solely by the angle $\theta$ in the Bloch sphere. All other state parameters remain unchanged during the experiment, ensuring that the magnitude of all overlaps -- and therefore two-photon exchange contributions -- are constant.

\section{Estimating additional exchange contributions}
\label{app:additionalExchangeContributions}
Now because $\braket{b}{d}\neq0$ there will be additional two- and three-photon exchange contributions to $P_{1111}$, as can be seen by comparing Fig.~\ref{fig:fig1}f(i) and (ii). In particular we have
\begin{equation}
\label{eqn:bdOverlap}
\begin{aligned}
\braket{b}{d}&=\braket{t_{2}}{t_{3}}\times\braket{+}{\theta}\\
&=\braket{t_{2}}{t_{3}}\times\frac{1}{2}(1+e^{i\theta})\\
&=\braket{t_{2}}{t_{3}}\cos(\theta/2) e^{i\theta/2}.
\end{aligned}
\end{equation}
This can lead to additional $\theta$-dependent variations so it important to investigate their contributions relative to the desired four-photon term. We use the expression for the general scattering probability from equation~(\ref{eqn:generalProb}) to write out all the exchange contributions to $P_{1111}$ for the experimentally realised input states of equation~(\ref{eqn:appStatePrep}) with Gaussian wavepackets and the interferometer in equation~(\ref{eqn:appQuitterUnitary}) (see Table~\ref{table:exchangeContributions}).
\begin{table}[ht]
	\caption{Exchange contributions to coincidence probability}
	\centering
	\begin{tabular}{c c c}
		\hline\hline
		Cycle $\rho$ & Matrix permanent & State dependence \\ [0.5ex]
		\hline
		$\mathbb{I}$ & $\frac{3}{32}$ & 1 \\
		(1,2) & $-\frac{1}{32}$ & $\mathbf{\abs{\braket{b}{d}}}^2$ \\
		(1,3) & $-\frac{1}{32}$ & $\abs{\braket{c}{d}}^{2}$ \\
		(1,4) & $-\frac{1}{32}$ & $\abs{\braket{a}{d}}^{2}$ \\
		(2,3) & $-\frac{1}{32}$ & $\abs{\braket{b}{c}}^{2}$ \\
		(2,4) & $-\frac{1}{32}$ & $\abs{\braket{a}{b}}^{2}$ \\
		(3,4) & $-\frac{1}{32}$ & $\mathbf{\abs{\braket{a}{c}}}^2$ \\
		(1,2,3) & $\frac{1}{32}$ &  $\mathbf{\braket{d}{b}}\braket{b}{c}\braket{c}{d}$\\ 
		(1,2,4) & $\frac{1}{32}$ & $\mathbf{\braket{d}{b}}\braket{b}{a}\braket{a}{d}$ \\
		(1,3,2) & $\frac{1}{32}$ & $\braket{d}{c}\braket{c}{b}\mathbf{\braket{b}{d}}$ \\
		(1,3,4) & $\frac{1}{32}$ & $\braket{d}{c}\mathbf{\braket{c}{a}}\braket{a}{d}$\\
		(1,4,2) & $\frac{1}{32}$ & $\braket{d}{a}\braket{a}{b}\mathbf{\braket{b}{d}}$ \\ [1ex]
		\hline
	\end{tabular}
	\quad
	\begin{tabular}{c c c}
		\hline\hline
		Cycle $\rho$ & Matrix permanent & State dependence \\ [0.5ex]
		\hline
		(1,4,3) & $\frac{1}{32}$ & $\braket{d}{a}\mathbf{\braket{a}{c}}\braket{c}{d}$ \\
		(2,3,4) & $\frac{1}{32}$ & $\braket{b}{c}\mathbf{\braket{c}{a}}\braket{a}{b}$ \\
		(2,4,3) & $\frac{1}{32}$ & $\braket{b}{a}\mathbf{\braket{a}{c}}\braket{c}{b}$ \\
		(1,2)(3,4) & $\frac{3}{32}$ & $\vert\mathbf{\braket{b}{d}}\vert^2\mathbf{\abs{\braket{a}{c}}}^2$ \\
		(1,3)(2,4) & $\frac{1}{32}(\cos2\chi+2)$ & $\abs{\braket{c}{d}}^{2}\abs{\braket{a}{b}}^{2}$ \\
		(1,4)(2,3) & $\frac{1}{32}(\cos2\chi+2)$ & $\abs{\braket{a}{d}}^{2}\abs{\braket{b}{c}}^{2}$ \\
		(1,2,3,4) & $-\frac{1}{32}$ &  $\mathbf{\braket{d}{b}}\braket{b}{c}\mathbf{\braket{c}{a}}\braket{a}{d}$\\
		(1,2,4,3) & $-\frac{1}{32}$ &  $\mathbf{\braket{d}{b}}\braket{b}{a}\mathbf{\braket{a}{c}}\braket{c}{d}$\\
		(1,3,2,4) & $\frac{1}{32}(\cos2\chi-2)$ &  $\braket{d}{c}\braket{c}{b}\braket{b}{a}\braket{a}{d}$\\
		(1,3,4,2) & $-\frac{1}{32}$ &  $\braket{d}{c}\mathbf{\braket{c}{a}}\braket{a}{b}\mathbf{\braket{b}{d}}$\\
		(1,4,2,3) & $\frac{1}{32}(\cos2\chi-2)$ &  $\braket{d}{a}\braket{a}{b}\braket{b}{c}\braket{c}{d}$\\
		(1,4,3,2) & $-\frac{1}{32}$ &  $\braket{d}{a}\mathbf{\braket{a}{c}}\braket{c}{b}\mathbf{\braket{b}{d}}$\\ [1ex]
		\hline
	\end{tabular}
	\label{table:exchangeContributions}
\end{table}

\begin{figure}[h]
	\centering
	\includegraphics[width=0.95\textwidth]{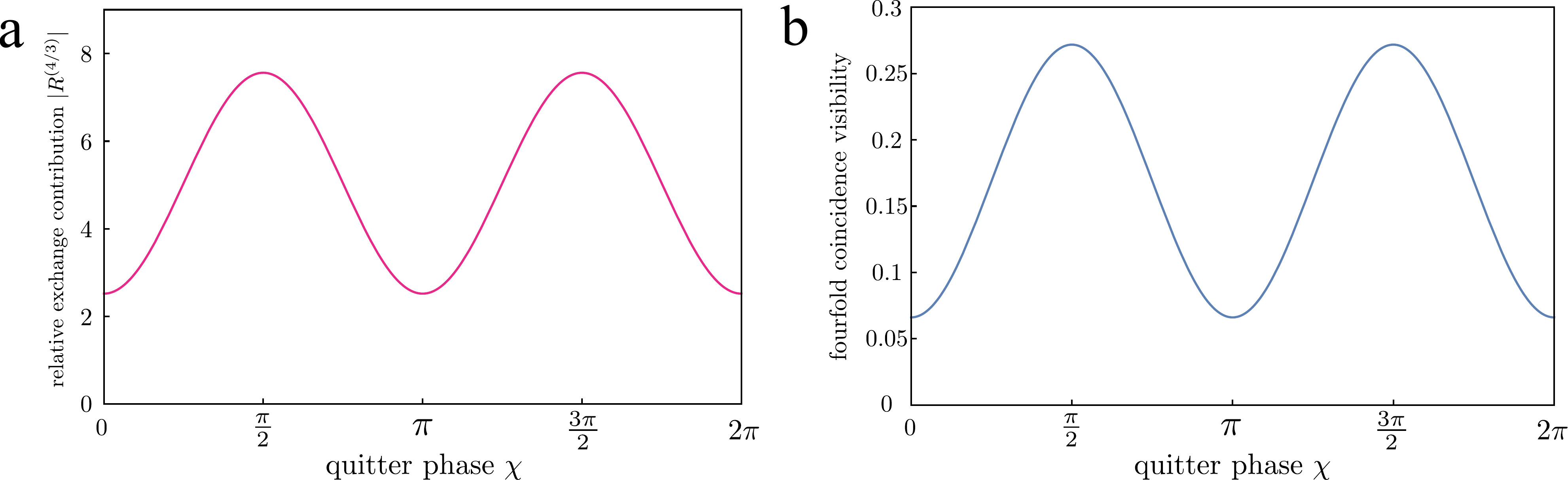}
	\caption{\label{fig:appExchangeAndVisWithPhase}\textbf{a} The absolute value of the ratio of the relative contributions of the three- and four-photon exchanges in equation~(\ref{eqn:contributionRatio}) as a function of the quitter phase $\chi$. \textbf{b} Four-photon signal visibility as a function of $\chi$, including additional exchange terms. }
\end{figure}

The sum of the products of all matrix permanents and corresponding state dependences gives the overall coincidence probability for the experiment. The emboldened scalar products are those we earlier assumed zero in order to eliminate the contributions from three-photon exchange, and ignoring these terms would recover the ideal $P_{1111}$ of equation~\ref{eqn:appP1111}. Omitting terms with $\braket{a}{c}=0$ (assuming orthogonal polarisations are possible in practice), we can sum terms to find the extra $\theta$-dependent contributions. The additional two-photon exchange contribution is:
\begin{equation}
\begin{aligned}
P^{(2)}&=-\frac{1}{32}\abs{\braket{b}{d}}^{2}\\
&=-\frac{1}{64}\abs{\braket{t_{2}}{t_{3}}}^{2}(1+\cos\theta),
\end{aligned}
\end{equation}
where we have substituted $\braket{b}{d}$ from equation~\ref{eqn:bdOverlap}. The extra three-photon exchange contributions can be simplified by setting $\abs{\braket{t_{1}}{t_{2}}}=\abs{\braket{t_{1}}{t_{3}}}=1/\sqrt{2}$ (a good approximation for the Gaussian wavepackets in Fig.~\ref{fig:temporalModes}b) and also substituting for $\braket{b}{d}$ to give:
\begin{equation}
\begin{aligned}
P^{(3)}&=\frac{1}{16}\mathrm{Re}[\braket{d}{b}\times(\braket{b}{c}\braket{c}{d}+\braket{b}{a}\braket{a}{d})],\\
&=\frac{1}{16}\mathrm{Re}[\abs{\braket{t_{2}}{t_{3}}}\cos(\theta/2) e^{-i\theta/2}\times\frac{1}{4}(e^{i\theta}+1)]\\
&=\frac{1}{64}\abs{\braket{t_{2}}{t_{3}}}(1+\cos\theta).
\end{aligned}
\end{equation}
For comparison we also write down the desired four-photon contribution:
\begin{equation}
\begin{aligned}
P^{(4)}&=\frac{1}{32}(\cos 2\chi-2)\times 2\mathrm{Re}[\braket{a}{b}\braket{b}{c}\braket{c}{d}\braket{d}{a}]\\
&=-\frac{1}{256}\times(2-\cos 2\chi)\times\cos\theta.
\end{aligned}
\end{equation}
Thus the presence of $\braket{b}{d}\neq0$ leads to an additional two-photon contribution $P^{(2)}$ that slightly \textit{increases} the expected $-\cos\theta$ variation, but also additional three-photon terms $P^{(3)}$ that \textit{decrease} this signal. The constant terms in $P^{(3)}$ and $P^{(4)}$ will decrease the $P_{1111}$ visibility slightly by adding a flat background. We now consider the relative size of the different $\cos\theta$ contributions from each type of exchange:
\begin{equation}
\begin{aligned}
\label{eqn:contributionRatio}
R^{(4/2)}&=\frac{(2-\cos 2\chi)}{4\abs{\braket{t_{2}}{t_{3}}}^{2}},\\
R^{(4/3)}&=-\frac{(2-\cos 2\chi)}{4\abs{\braket{t_{2}}{t_{3}}}}.
\end{aligned}
\end{equation}
We aim to experimentally set $\abs{\braket{t_{2}}{t_{3}}}=0.1$ and so the two-photon contribution, that goes as the square of the overlap magnitude, will be much smaller than the three-photon contribution. We plot the magnitude of $R^{(4/3)}$ against $\chi$ in Fig.~\ref{fig:appExchangeAndVisWithPhase}a. It reaches a maximum of 7.6 and a minimum of 2.5 and, given uniformly distributed $\chi$, averages to $\sim$5. During the experiment we record the true $\chi$ values sampled and, thanks to the locking procedure, this slightly increases the average relative contribution so that the fourfold term is six times greater than the threefold one.

The revised fourfold coincidence probability can be found by inserting the various $r_{ij}$ into expressions in Table~\ref{table:exchangeContributions}, and the visibility is given by $\nicefrac{(\mathrm{max}-\mathrm{min})}{\mathrm{max}}$, where the maximum probability occurs when $\theta=\pi$ and the minimum when $\theta=0,2\pi$. The visibility is plotted against $\chi$ in Fig.~\ref{fig:appExchangeAndVisWithPhase}b.

We earlier saw that for ideal pairwise distinguishable states, the visibility of $P_{1111}$ achieves a maximum of $30\%$ for $\chi=(2n+1)\pi/2$ and a minimum of $10\%$ for $\chi=n\pi$. For our Gaussian wavepackets, these visibilities are reduced respectively to $27.2\%$ and $6.6\%$ due to the extra three-photon terms. Uniformly distributed $\chi$ would lead to an average visibility of $16.9\%$ but using the actual $\chi$ values measured during the experiment increases this to a predicted $21.0\%$. Importantly the majority of any observed $-\cos\theta$ variation derives from the desired four-photon interference since two-photon terms are smaller by a factor of $\sim\abs{\braket{t_{2}}{t_{3}}}^{2}\ll 1$ and any three-photon terms give $+\cos\theta$ variation.

\section{Subtracting statistics from double emissions}
\label{app:subtractingStatistics}
\begin{figure}[h]
\centering
\includegraphics[width=0.85\textwidth]{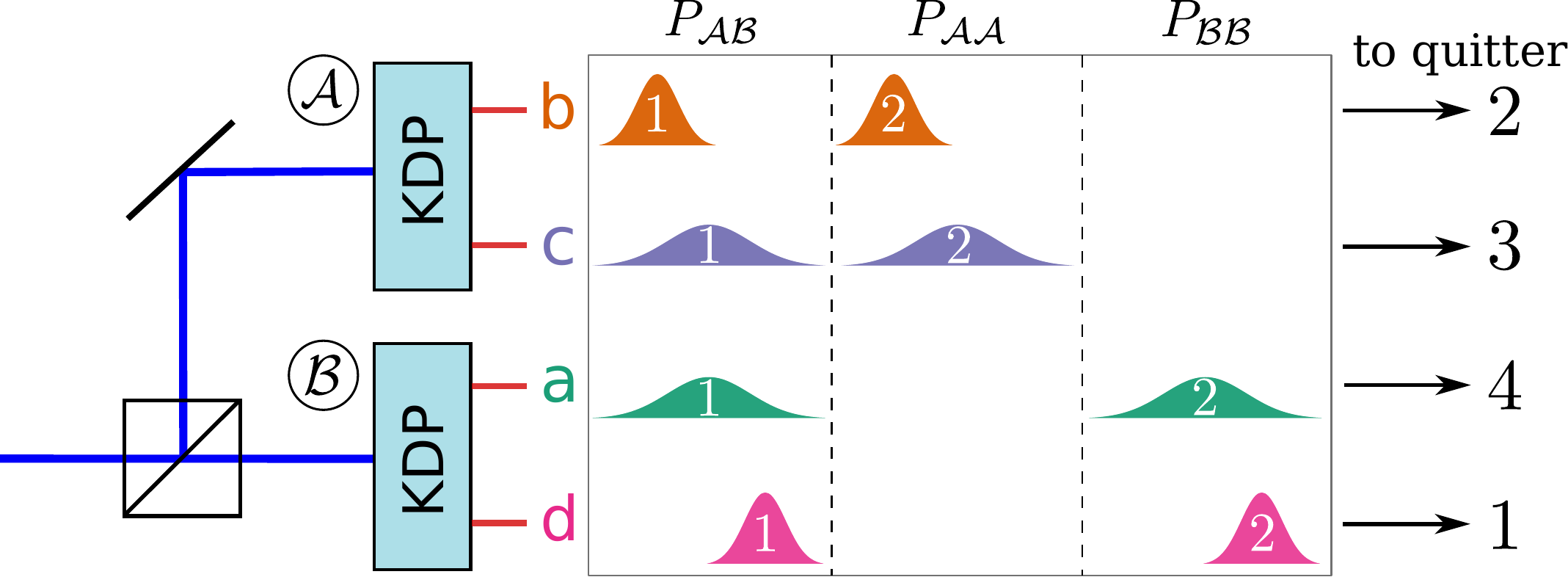}
\caption{\label{fig:appSourceInterference}Configuration of our two SPDC sources (labelled $\mathcal{A}$ and $\mathcal{B}$) to generate the states in equation~(\ref{eqn:appStatePrep}), and with the associated quitter input ports labelled from equation~(\ref{eqn:stateOrder}). The inset box shows the possible emissions of four photons from the sources, with the number of photons in each state labelled and with $P_{\mathcal{AB}},P_{\mathcal{AA}},P_{\mathcal{BB}}$ as the preparation probabilities.}
\end{figure}

In Fig.~\ref{fig:appSourceInterference} we show the preparation of the four photons and their injection order into the quitter interferometer. This particular input port configuration both maximises the four-photon signal visibility and also allows monitoring of the quitter's internal phase $\chi$ by blocking source $\mathcal{B}$ and using the technique shown in Fig.~\ref{fig:lockingDemo}.

The state generated by a factorable spontaneous parametric down-conversion source is a two-mode squeezed vacuum:
\begin{equation}
\label{eqn:TMSV}
\ket{\Psi_{\mathrm{TMSV}}}=\sqrt{1-\lambda^2}\sum_{n=0}^{\infty}\lambda^{n}\ket{n_{s},n_{i}}.
\end{equation}
$\lambda$ is the squeezing parameter, \textit{s} and \textit{i} denote the signal and idler modes respectively, and $n$ is the occupation of that mode. In our experiment we generate two such states, inject them into a quitter, and perform coincidence counting (Fig.~\ref{fig:appSourceInterference}). The quitter erases spatial information about which source fired: hence it is impossible to know which source or sources fired to generate the four photons (i.e. source $\mathcal{A}$ or $\mathcal{B}$ fired twice, or each source fired once as depicted in Fig.~\ref{fig:appSourceInterference}). As a result, there are additional interference effects arising due to coherences with the vacuum components of each state~\cite{Ou1989,Zou1991}. This was also noted in~\cite{Carolan2014} where a similar method was used to generate four photons; they applied a Pancharatnam (Berry) phase using waveplates on one of the modes to remove these coherences. We use this same technique so that, on averaging data acquired at four different Pancharatnam phases, the effective input state is:
\begin{equation}
\rho_{in}^{eff}=\mathcal{N}\left(P_{\mathcal{AB}}\ket{\Psi_{\mathcal{AB}}}\bra{\Psi_{\mathcal{AB}}}+P_{\mathcal{AA}}\ket{\Psi_{\mathcal{AA}}}\bra{\Psi_{\mathcal{AA}}}+P_{\mathcal{BB}}\ket{\Psi_{\mathcal{BB}}}\bra{\Psi_{\mathcal{BB}}}\right),
\end{equation}
where $\mathcal{N}$ is a normalisation factor to ensure the density matrix has unit trace, and we have allowed for the preparation probabilities $P_{\mathcal{AB}},P_{\mathcal{AA}},P_{\mathcal{BB}}$ to be different. The first term corresponds to each source firing once and means successful preparation of the desired input state $\ket{\Psi_{\mathcal{AB}}}=\ket{1_{\ket{d}},1_{\ket{b}},1_{\ket{c}},1_{\ket{a}}}$, where ordering in the ket corresponds to quitter input port. The other two terms are cases of probabilistically preparing the states $\ket{\Psi_{\mathcal{AA}}}=\ket{0,2_{\ket{b}},2_{\ket{c}},0}$ and $\ket{\Psi_{\mathcal{BB}}}=\ket{2_{\ket{d}},0,0,2_{\ket{a}}}$. The overall fourfold coincidence probability includes contributions from all three terms:
\begin{equation}
\label{eqn:totalP1111signal}
P_{1111}^{tot}=\mathcal{N}(P_{\mathcal{AB}}\times P_{1111}^{\mathcal{AB}}+P_{\mathcal{AA}}\times P_{1111}^{\mathcal{AA}}+P_{\mathcal{BB}}\times P_{1111}^{\mathcal{BB}}).
\end{equation}
The probability $P_{1111}^{\mathcal{AB}}=P_{1111}$ from equation~\ref{eqn:P1111} and contains the $\varphi_{abcd}$ term we want. The coincidences due to single sources firing twice depend only on pairwise distinguishabilities $r_{ij}^{2}$:
\begin{equation}
\label{eqn:doubleEmissionCoincidences}
\begin{aligned}
P_{1111}^{\mathcal{AA}}&=\frac{1}{32}\left(3-4r_{bc}^{2}+r_{bc}^{4}(2+\cos 2\chi)\right),\\
P_{1111}^{\mathcal{BB}}&=\frac{1}{32}\left(3-4r_{ad}^{2}+r_{ad}^{4}(2+\cos 2\chi)\right).
\end{aligned}
\end{equation}
For our state preparation $r_{bc}=r_{ad}$, so these scattering probabilities should be equal. On averaging over the sampled $\chi$ values, these terms contribute a flat background to $P_{1111}^{tot}$. The preparation probabilities are determined by the sources' squeezing parameters and input losses. For balanced pumping power and no losses they would all be given by the firing probability $\lambda^{4}\left(1-\lambda^{2}\right)^{2}$ and $P_{1111}^{tot}$ would be given by a balanced mixture of the scattering probabilities in equation~\ref{eqn:totalP1111signal}. The effect of the flat background contributions to $P_{1111}^{tot}$ would be to decrease the visibility of the signal we are after (isolated $P_{1111}^{\mathcal{AB}}$) from $21.0\%$ to $7.8\%$, given the sampled $\chi$ values.

During the experiment we separately record the double emission fourfolds by blocking each source in turn using shutters. It is important to note that this changes the probability of one source not firing from $(1-\lambda^{2})$ to 1. Hence the measured backgrounds are 
\begin{equation}
P_{\mathcal{AA}}^{bg}=\frac{P_{\mathcal{AA}}}{(1-\lambda^{2})}\times P_{1111}^{\mathcal{AA}},\qquad
P_{\mathcal{BB}}^{bg}=\frac{P_{\mathcal{BB}}}{(1-\lambda^{2})}\times P_{1111}^{\mathcal{BB}}.
\end{equation}
Correcting for this factor (which is close to unity because for these sources $\lambda=0.16$), we then subtract the independently measured backgrounds from $P_{1111}^{tot}$ to leave the desired $P_{1111}^{\mathcal{AB}}$.

Whilst the pumping powers for each source were equal, the input losses to the interferometer were slightly imbalanced meaning $P_{\mathcal{BB}}\approx 1.24\times P_{\mathcal{AA}}$ (the mean total counts per point in Fig.~\ref{fig:fig4}e are 1,740 and 2,160 for $\mathcal{AA}$ and $\mathcal{BB}$ respectively). We account for this by including input transmissions $0\leq\eta_{i}\leq1,i\in(1,2,3,4)$ and rewriting the total fourfold coincidence probability as
\begin{equation}
\label{eqn:Ptot}
P_{1111}^{tot'}=\mathcal{N}'(\eta_{1}\eta_{2}\eta_{3}\eta_{4}\times P_{1111}^{\mathcal{AB}}+\eta_{2}^{2}\eta_{3}^{2}\times P_{1111}^{\mathcal{AA}}+\eta_{1}^{2}\eta_{4}^{2}\times P_{1111}^{\mathcal{BB}}).
\end{equation}
We have factored out the common source firing probability and included input transmissions explicitly. From the total recorded counts when blocking individual sources, we estimate $\eta_{1}\eta_{4}/\eta_{2}\eta_{3}\approx\sqrt{1.24}=1.1$. This leads us to a revised total fourfold signal
\begin{equation}
\label{eqn:P1111tot2}
P_{1111}^{tot'}=\mathcal{N''}(1.11\times P_{1111}^{\mathcal{AB}}+P_{1111}^{\mathcal{AA}}+1.24\times P_{1111}^{\mathcal{BB}}).
\end{equation}
This can be compared to equation~\ref{eqn:totalP1111signal} where $P_{\mathcal{AB}}=P_{\mathcal{AA}}=P_{\mathcal{BB}}$ for equal preparation probabilities. The revised preparation probabilities have a very small effect, essentially leaving the total signal visibility unchanged at $7.8\%$ Performing the same background subtraction just mentioned, the desired $P_{1111}^{\mathcal{AB}}$ signal's visibility is unaffected by imbalanced preparation probabilities because you effectively postselect on the required $\mathcal{AB}$ emission event.

Six-photon emissions are also possible from this pair of sources with a probability $\lambda^{2}\approx 0.16^2=2.6\%$ of that for four-photon emission. The possible firings are $\mathcal{AAA,BBB,AAB,ABB}$ and they can all lead to quitter outputs that register as fourfold coincidences. The first two depend only on pairwise distinguishabilities and again average to a flat background for our sampled $\chi$ values. Furthermore these are removed by the background subtraction just described. The other two possible firings -- where one source fires twice and the other once -- cannot be measured separately but each contain a constant term and a small $-\cos\varphi_{abcd}$ term. The constant term dominates and so these sixfold terms decrease the fourfold signal visibility. Including these higher-order emissions only very slightly reduces the total signal visibility, and that of the background-subtracted $P_{1111}^{\mathcal{AB}}$ signal by about $1\%$. Emissions of eight or more photons are very rare so we ignore their effect on the fourfold signals.

From independent HOM measurements we infer there is a small amount of unintended distinguishability between each pair, most likely arising from slight spectral mismatch. We model this by including an additional factor in the pairwise distinguishabilities of $r_{ij}^{2}=0.95$. This reduces the visibility of the total fourfold signal $P_{1111}^{tot'}$ to a predicted $6.7\%$, and the background-subtracted $P_{1111}$ to $17.9\%$.

\section{Preparing temporal and polarisation modes}
\label{app:exptPreparation}
\subsection{Walking off wavepackets temporally}
\label{app:tempWalkoff}
\begin{figure}[h]
	\centering
	\includegraphics[width=0.625\textwidth]{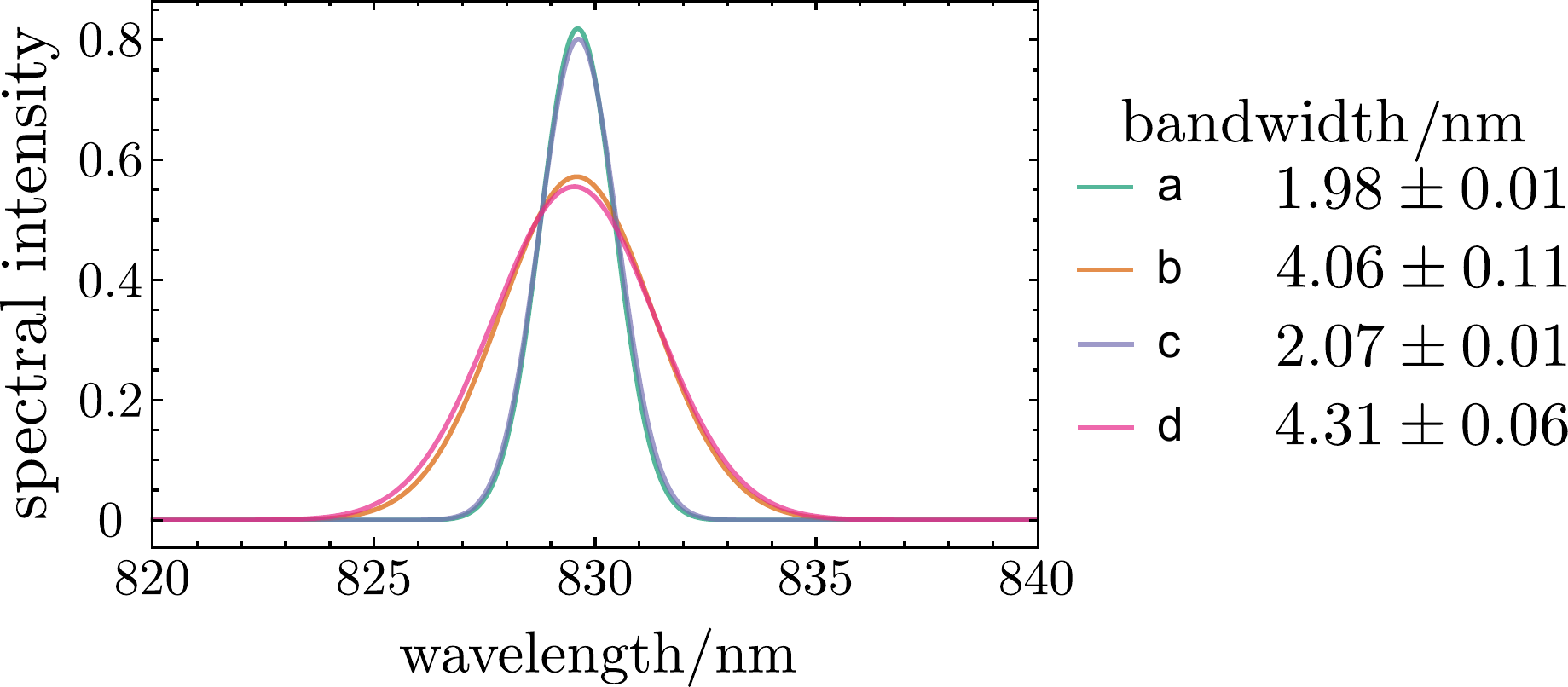}
	\caption{\label{fig:fourPhotSpectra} We insert 3~nm filters on the narrowband signal modes to achieve filtered bandwidths of 2~nm for the states labelled $a,c$. For the broadband idler modes, a pair of bandpass filters are applied to trim each side of the spectrum and give $\sim4.2$~nm bandwidths for the states labelled $b,d$.}
\end{figure}

The bandwidths of the spectrally unentangled and mean-wavelength degenerate signal and idler photons generated by the KDP SPDC sources are 2.5~nm and 12.5~nm respectively, meaning the temporal duration of the signal photon's wavepacket is five times that of the idler's. We saw in Supplementary~\ref{app:distStatePrep} that the desired ratio is around $2.1:1$ and so we apply spectral filtering (see Fig.~\ref{fig:fourPhotSpectra}).

The positions of delay stages that correspond to simultaneous arrival of all four photons at the quitter are found by measuring combinations of HOM dips between different arms of the SPDC sources. Comparison of the dip widths and measured spectra confirms transform-limited wavepackets.

\begin{figure}[ht]
	\centering
	\includegraphics[width=0.45\textwidth]{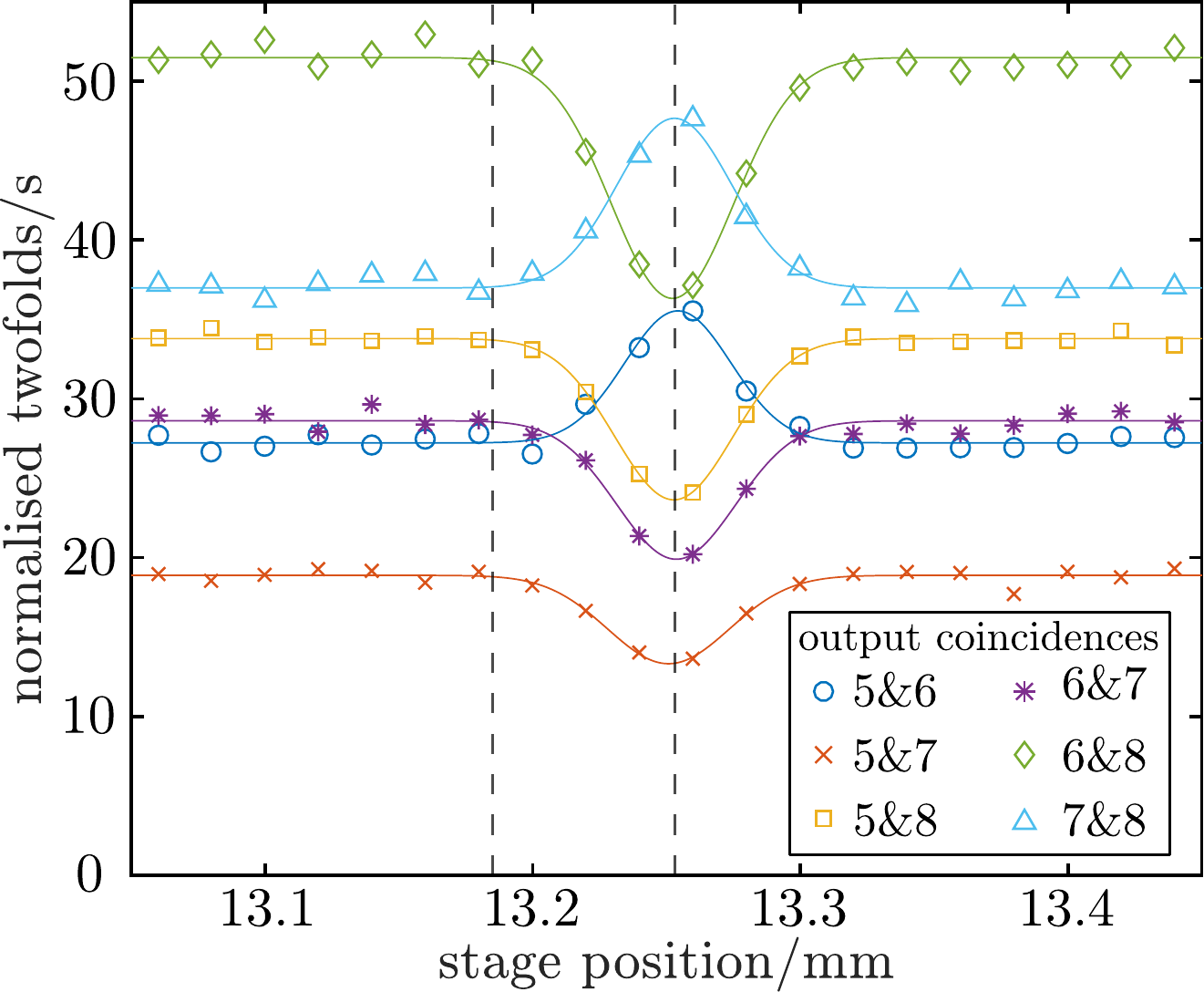}
	\caption{\label{fig:walkOffReference}Two-photon interference between the temporally narrow independent thermal states when they have matched polarisations, giving dips with $\sim33\%$ visibility and peaks with $\sim25\%$ visibility. We set the delay stage to the position indicated by the dashed line on the left for a temporal overlap of 0.1.}
\end{figure}

To achieve the distinguishability of temporal modes so $\braket{b}{d}\approx0$, the states' polarisations are matched and the two-photon interference for thermal states is recorded as the relative delay is varied (see Fig.~\ref{fig:walkOffReference}). We set $\abs{\braket{t_{2}}{t_{3}}}\leq0.1$ by adjusting the delay stage to the position indicated by the dashed line on the left. The arrival time $t_{1}$ of states $\ket{a}$ and $\ket{c}$ is adjusted to achieve the configuration shown in Fig.~\ref{fig:temporalModes}b where $\abs{\braket{t_{1}}{t_{2}}}=\abs{\braket{t_{1}}{t_{3}}}\approx1/\sqrt{2}$.

\subsection{Calibrating waveplate angle and $\varphi_{abcd}$}
\label{app:calibrateWPphase}
 On rotation of the state $\ket{c}$ to vertical polarisation, the absence of a HOM dip with the horizontally polarised state $\ket{a}$ verifies the required distinguishability $\braket{a}{c}=0$. To prepare state $\ket{b}$ in $\ket{+}$, we first orient the polarisation of $\ket{d}$ somewhere in the equator of the Bloch sphere and calibrate an output polarisation analyser so this is the system's diagonal polarisation state. Now we send in $\ket{b}$ and manipulate compensation waveplates to match this definition of diagonal. Since it is only relative orientation of these states in the equator that matters, this corresponds to $\varphi_{abcd}=0$. 
 
\begin{figure}[h!]
	\centering
	\includegraphics[width=0.68\textwidth]{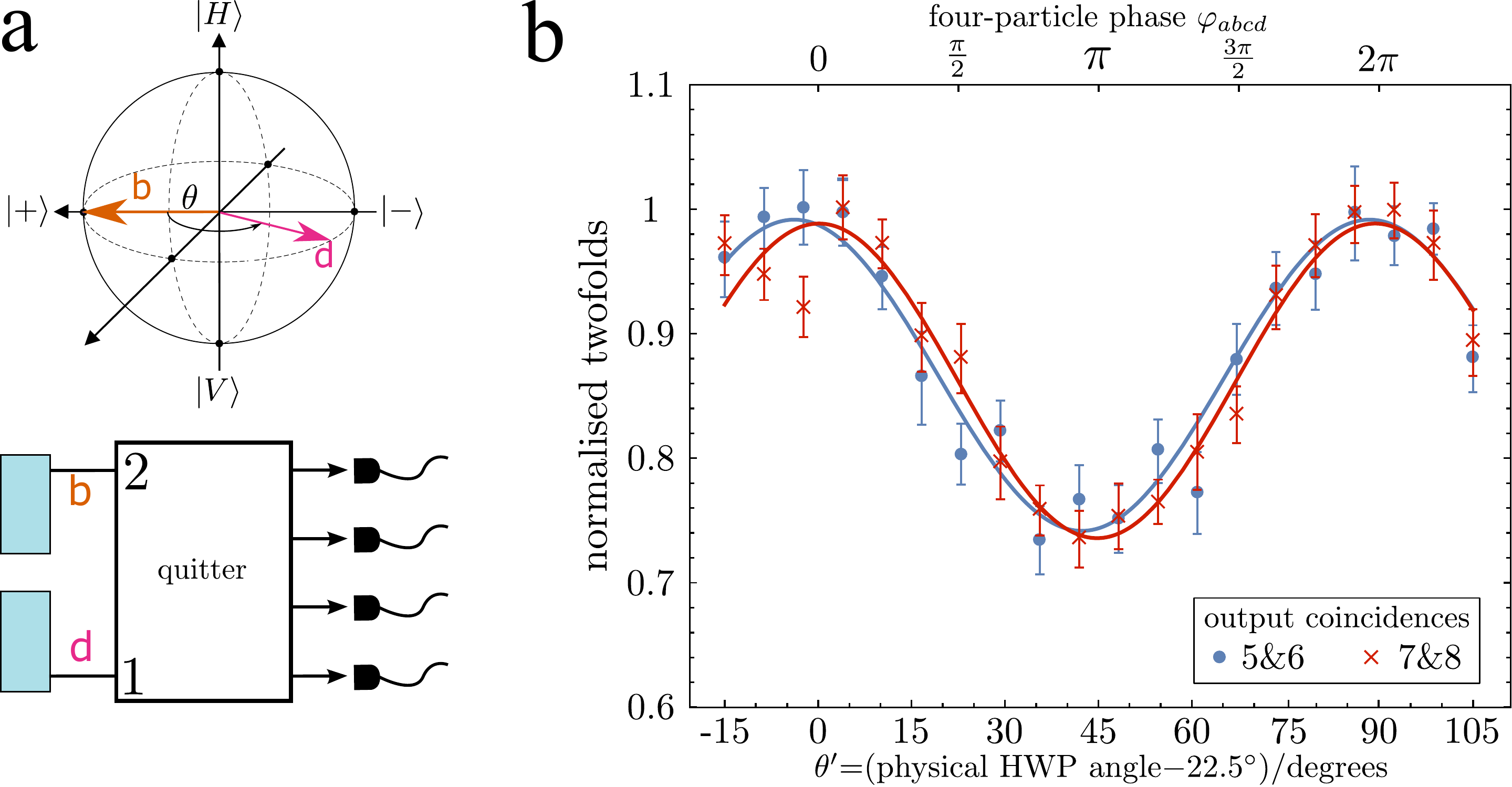}
	\caption{\label{fig:calibratingPhaseWP} \textbf{a} In order to calibrate the waveplate angle to the four-particle phase $\varphi_{abcd}$, we temporally align the states labelled $b$ and $d$ and rotate the latter in the equator of the Bloch sphere whilst recording twofolds at the quitter outputs. \textbf{b} This set of output coincidences corresponds to the sampling of bunching on a balanced beam splitter of two photons from independent sources (see Fig~\ref{fig:appFoldedQuitter}a). Indistinguishability is achieved when $4\theta'=\varphi_{abcd}=0$ and the shape is $\sim\cos^{2}2\theta'$. These thermal state signals are predicted to have $25\%$ visibility. The shape of these sets of coincidences are used to calibrate the physical waveplate angle through $\theta'=\theta/4$ with the four-particle phase $\varphi_{abcd}$.}
\end{figure}

 A quarter- and a half-wave plate are used to prepare the polarisation of $\ket{d}$:
 \begin{equation}
 U_{QWP}\left(\frac{\pi}{4}\right)\cdot U_{HWP}\left(\theta'+\frac{\pi}{8}\right)\ket{H}=\frac{1}{\sqrt{2}}\left(\ket{H}+e^{4i\theta'}\ket{V} \right),
 \end{equation}
 where we have inserted the physical angles of the waveplates in radians. Rotating the angle $\theta'$ from 0 to $\pi/2$ will change the angle in the Bloch sphere $\theta=4\theta'$ from 0 to $2\pi$. In order to calibrate $\theta'$ to the corresponding $\varphi_{abcd}$, we temporally overlap the spectrally indistinguishable states $\ket{b},\ket{d}$ and rotate the HWP whilst monitoring twofolds at the quitter outputs (see Fig.~\ref{fig:calibratingPhaseWP}). These unheralded thermal state interference signals are used to associate the physical waveplate angle with the four-particle phase $\varphi_{abcd}$. 

\section{Details of analysis and additional counting statistics}
\label{appxSec:dataNorm}
\subsection{Estimating contributions from lower-order exchanges}
During each run of the experiment (where $\varphi_{abcd}$ is varied across the range shown in Fig.~\ref{fig:fig4}), we use shutters to block different arms of the SPDC sources to allow independent measurement of variations in lower-order exchange contributions for each $\varphi_{abcd}$ step. Many such `sweeps' are recorded and analysed to determine the average variations in the different counting channels. In the following we describe how these statistics are used to assess the impact of observed variations in one-, two- and three-photon statistics on the recorded fourfold signals.
\subsubsection{Single source arm open}
\label{appxsec:singleArms}
Due to changes in laboratory conditions, the coupling to the quitter varies slightly between sweeps. However these changes are slow compared to the time taken to acquire data for a single sweep: the total singles counts recorded at the quitter outputs (when injecting all source arms) vary by a maximum of around $5\%$ over more than five hours, whilst a single sweep of $\varphi_{abcd}$ (where at each setting data is recorded for all different shutter configurations) takes approximately three minutes. We can therefore assume that the coupling is constant for each individual sweep of $\varphi_{abcd}$.

We care about changes in counting signals that are independent of these slow coupling variations between sweeps. It is useful to cast statistics in terms of actual counts instead of just probabilities. We define the true number of singles and twofolds from a single source as
\begin{equation}
\label{eqn:cCounts}
R_{\tau}=\lambda^{2}(1-\lambda^{2})\times S\times \tau,
\end{equation}
where the first term is the probability of successful emission per trial (from equation~\ref{eqn:TMSV}), $S$ is the laser's repetition rate (equivalent to the rate of trials) and $\tau$ is the associated measurement time in seconds. We also denote the transmission on input $i\in(1,2,3,4)$ by $0\leq\eta_{i}\leq1$ and the transmission on output $j\in(5,6,7,8)$ by $0\leq\eta_{j}\leq1$. These can vary slowly between sweeps and result in changes to the singles counts.

\begin{figure}[ht]
	\centering
	\includegraphics[width=0.8\textwidth]{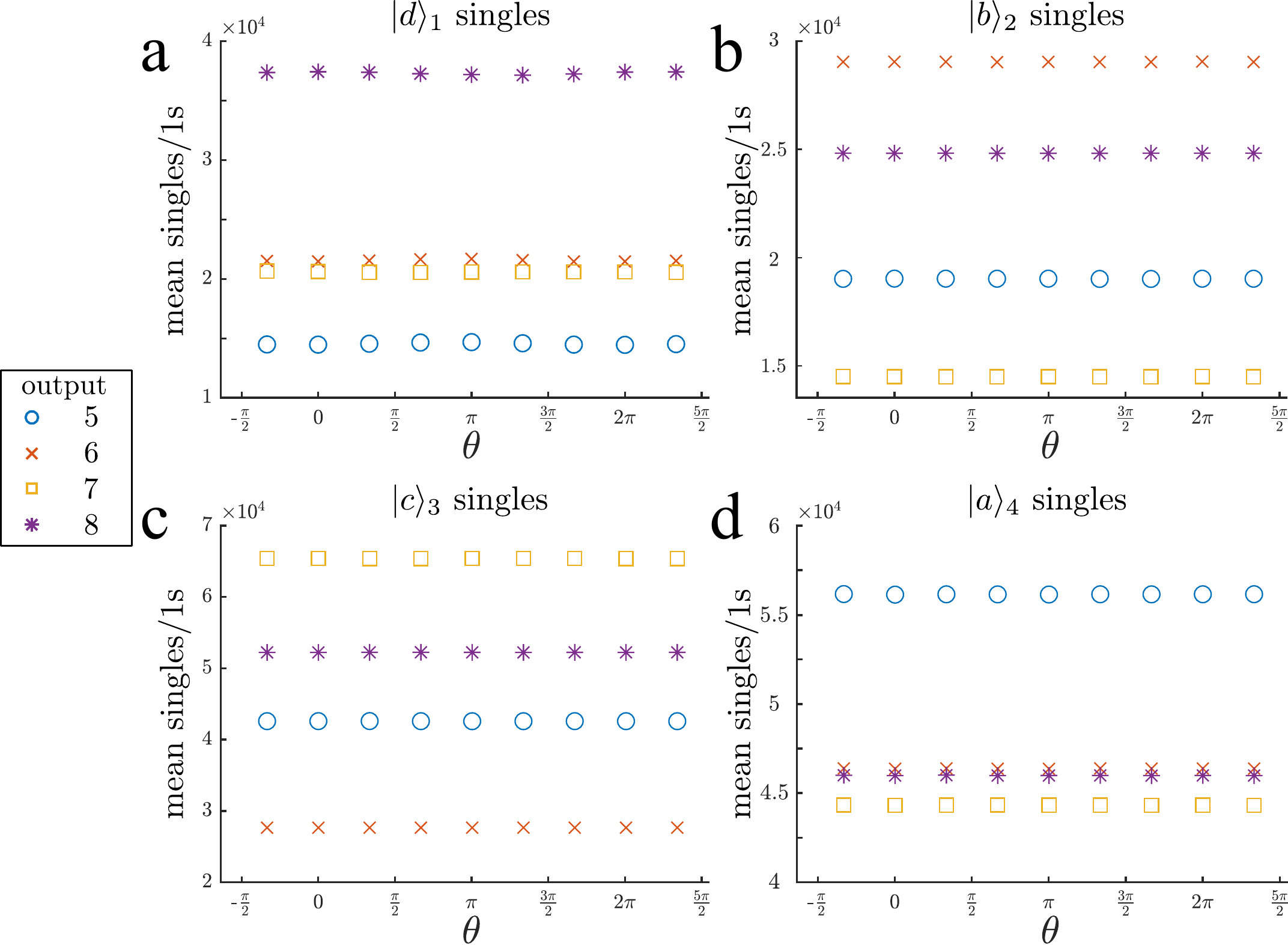}
	\caption{\label{fig:appSinglesCounts}Output singles for each individual quitter input recorded by blocking other arms. The input port is denoted by the state's subscript and error bars are smaller than markers so are omitted. See main text for details.}
\end{figure}

For each sweep of $\varphi_{abcd}$ we use nine different values of the angle $\theta$ in the Bloch sphere, equally spaced from $-\pi/3$ to $7\pi/3$. In contrast to Fig.~\ref{fig:fig4} in the main text where the x-axes are labelled by the four-particle phase $\varphi_{abcd}$, in this Supplementary we use $\theta$ as our label to highlight the experimentally varied polarisation angle of state $\ket{d}$. The angles take the same value, as shown in Supplementary~\ref{app:calibrateWPphase}, but have different interpretations: one is an angle in the Bloch sphere, and the other is a multiparticle phase. We define $C^{i}_{j}(\theta)$ as the singles recorded at output $j$ when the state at input $i$ is injected into the quitter, and any dependence on $\theta$ arises only when $i=1$. These singles counts are
\begin{equation}
C^{i}_{j}(\theta)=R_{\tau}\times\eta_{i}\eta_{j}\times P^{i}_{j}(\theta),
\end{equation}
where $P^{i}_{j}(\theta)$ is the probability of a single photon scattering from input $i$ to output $j$ in the quitter, and  is independent of $\chi$. This probability can depend on the polarisation angle $\theta$ when $i=1$: we found that if the polarisation axes were not aligned with the beam splitter cube's axes and $\ket{d}$ is rotated in the equator of the Bloch sphere, then the total number of transmitted singles is constant but those on individual outputs can vary. Changes in counts on outputs 5 and 6 were correlated with each other, but anti-correlated with those on outputs 7 and 8. We therefore aligned the polarisation axes to the beam splitter cube's axes to minimise these variations but still retain a possible $\theta$ dependence. For each sweep we consider normalising the singles counts at each $\theta$ step to the average over the sampled $\theta$, so
\begin{equation}
\label{eqn:appxSinglesNorm}
\bar{C}^{i}_{j}(\theta)=\frac{C^{i}_{j}(\theta)}{\frac{1}{9}\sum_{\theta}C^{i}_{j}(\theta)}.
\end{equation}
This cancels out the effect of slow variations in the input and output transmissions, and in the generation rate $R_{\tau}$, also assumed constant over each sweep. Deviation from unity would indicate that the single-particle scattering probability depends on $\theta$, and this could only be the case for $i=1$. We perform this normalisation for each individual sweep, and then average over all sweeps to recover the overall average variation for each singles channel, independent of the absolute count rate. We also calculate the standard error on the mean value of $\bar{C}^{i}_{j}(\theta)$ for each $\theta$ value sampled. In order to give an idea of count rates during the experiment, we then multiply these normalised statistics by the mean count rate at $\theta=-\pi/3$. In other words, we plot what the singles signals would look like if there were no slow drifts in coupling. This does not change the visibility of any variations, which is the critical parameter being investigated here. Results are shown in Fig.~\ref{fig:appSinglesCounts}.

Only the state $\ket{d}$ is varied during the experiment: its polarisation rotates in the Bloch sphere to change the four-particle phase $\varphi_{abcd}$ (Fig.~\ref{fig:calibratingPhaseWP}a). The singles at the quitter outputs when injecting only this state are shown in Fig.~\ref{fig:appSinglesCounts}a. Those at outputs 5 and 6 exhibit small structured variations of $1.35\%$ and $1\%$ respectively with $-\cos\theta$ shape, and those at output 8 have $0.7\%$ variation with $+\cos\theta$ shape, for an overall average of $\sim0.4\%$ fluctuation. We attribute these variations to a very small residual polarisation-dependence of the beam splitter in the quitter that we have not entirely eliminated: there is still a small mismatch of the polarisation axes and the beam splitter cube's axes. The output singles for the other three input states ($i=2,3,4$) are shown in Fig.~\ref{fig:appSinglesCounts}b-d and, since they are not adjusted during the experiment, are all constant within the error bars.
\subsubsection{Two source arms open}
\label{appxsec:twoArms}
Due to the variations in interferometer coupling, the absolute twofold rates also vary slowly over the course of the experiment. We would like to assess how constant the pairwise distinguishabilities $r_{ij}^{2}$ are throughout the experiment and so must remove the effect of underlying singles variations. In this section we show this can be accomplished by normalising the twofold counts to the product of the separately measured singles.

The rates for two photons produced from the same SPDC crystal (single source `$ss$', when $i,j=(1,4)$ or $(2,3)$) are proportional to $\lambda^2(1-\lambda^2)$, but the rates for two photons generated when one arm is used from each source (two sources `$ts$', for all other pairs of input ports) are instead proportional to $\lambda^4(1-\lambda^2)^2$. We denote the number of twofold counts recorded when opening inputs $i,j$ and measuring at outputs $k,l$ as:
\begin{equation}
\begin{aligned}
C^{i,j, (ss)}_{k,l}(r_{ij}^{2},\theta,\chi)&=R_{\tau}\times\eta_{i}\eta_{j}\eta_{k}\eta_{l}\times  P^{i,j}_{k,l}(r_{ij}^{2},\theta,\chi),\\
C^{i,j,(ts)}_{k,l}(r_{ij}^{2},\theta,\chi)&=R'_{\tau}\times\eta_{k}\eta_{l}\left(\eta_{i}\eta_{j}\times P^{i,j}_{k,l}(r_{ij}^{2},\theta,\chi)+\eta_{i}^{2}\times P^{i,i}_{k,l}(\theta)+\eta_{j}^{2}\times P^{j,j}_{k,l}(\theta)\right).
\end{aligned}
\end{equation}
$R'_{\tau}=R_{\tau}\times\lambda^{2}(1-\lambda^2)$ and $P^{i,j}_{k,l}(r_{ij}^{2},\theta,\chi)$ is the two-photon scattering probability for these inputs and outputs that depends on the pairwise distinguishability $r_{ij}^{2}$ of the states in inputs $i,j$ and, depending on the input ports, can also depend on $\theta$ and $\chi$ (see for example equation~\ref{eqn:phaseDepHOM}). For the two source case $ts$ we have included terms corresponding to each source firing twice that also lead to twofold coincidences at outputs $k,l$ by purely probabilistic scattering: there is no $\chi$ dependence and $i=j$ means the input photons are indistinguishable, so the $r_{ij}^{2}$ dependence is omitted, but possible dependence on $\theta$ (through polarisation dependence of the single-particle scattering probability if $i=1$ or $j=1$) is retained. In a similar way to the background subtraction mentioned in Supplementary~\ref{app:subtractingStatistics}, we separately measure the twofolds arising from a single source firing twice when only one arm is open and, with appropriate factors to accommodate different firing probabilities, subtract these counts from $C^{i,j,(ts)}_{k,l}(r_{ij}^{2},\theta,\chi)$ to find
\begin{equation}
\label{eqn:appxTwoSourceNorm}
C^{i,j,(ts)}_{k,l,(bgsub)}(r_{ij}^{2},\theta,\chi)=R'_{\tau}\times\eta_{i}\eta_{j}\eta_{k}\eta_{l}\times  P^{i,j}_{k,l}(r_{ij}^{2},\theta,\chi).
\end{equation}

We now proceed to divide these twofolds by the product of the relevant separately measured singles to give
\begin{equation}
\label{eqn:twofoldsNorm}
\begin{aligned}
\tilde{C}^{i,j,(ss)}_{k,l}(r_{ij}^{2},\theta,\chi)&=\frac{C^{i,j,(ss)}_{k,l}(r_{ij}^{2},\theta,\chi)}{C^{i}_{k}(\theta)\times C^{j}_{l}(\theta)}=\frac{R_{\tau}}{R^{2}_{\tau}\times P^{i}_{k}(\theta) P^{j}_{l}(\theta)}\times P^{i,j}_{k,l}(r_{ij}^{2},\theta,\chi),\\
\tilde{C}^{i,j,(ts)}_{k,l,(bgsub)}(r_{ij}^{2},\theta,\chi)&=\frac{C^{i,j,(ts)}_{k,l,(bgsub)}(r_{ij}^{2},\theta,\chi)}{C^{i}_{k}(\theta)\times C^{j}_{l}(\theta)}=\frac{R'_{\tau}}{R^{2}_{\tau}\times P^{i}_{k}(\theta) P^{j}_{l}(\theta)}\times P^{i,j}_{k,l}(r_{ij}^{2},\theta,\chi).
\end{aligned}
\end{equation}
This procedure removes the effect of slowly varying losses at the inputs and outputs. The constants $R_{\tau}$ and $R'_{\tau}$ could be calculated using known $\tau$ and squeezing parameter $\lambda$. The next step could then be to correct for the single-particle scattering probabilities to obtain the true two-photon scattering probabilities. However the single-particle scattering probabilities are not known to good accuracy because of sensitivities to losses, so we do not back out the two-photon probabilities.

Instead we perform the above procedure to find $\tilde{C}$ for each separate sweep of $\theta$ and then divide each sweep by its mean value (analogous to equation~\ref{eqn:appxSinglesNorm}) to cancel the effect of any slow variations in transmissions and generation rate. An important assumption is that this isolates $\theta$-dependent variations in the two-photon scattering probabilities. In the previous section we saw that the single-particle scattering probabilities for $\ket{d}$ on input port 1 have a small dependence on $\theta$. We therefore estimate the effect of averaging $1/P^{1}_{k}(\theta)\sim 1/(1+\frac{\varepsilon_{k}}{2}\cos\theta)$ over the nine sampled $\theta$ values, where $\varepsilon_{k}$ is the variation of counts on output $k$ (for example $\varepsilon_{5}=-0.0135$). This number differs from unity by at most only $0.1\%$ and so variations in the denominators of equations~\ref{eqn:twofoldsNorm} associated with changes in the single-particle scattering probabilities are negligible. Dividing the calculated $\tilde{C}$ by the mean for each sweep of $\theta$ therefore isolates the $\theta$ dependence of the two-photon scattering probability, independent of the single-particle scattering probabilities and the slowly varying transmissions and generation rate.

\begin{figure}[h]
	\centering
	\includegraphics[width=1\textwidth]{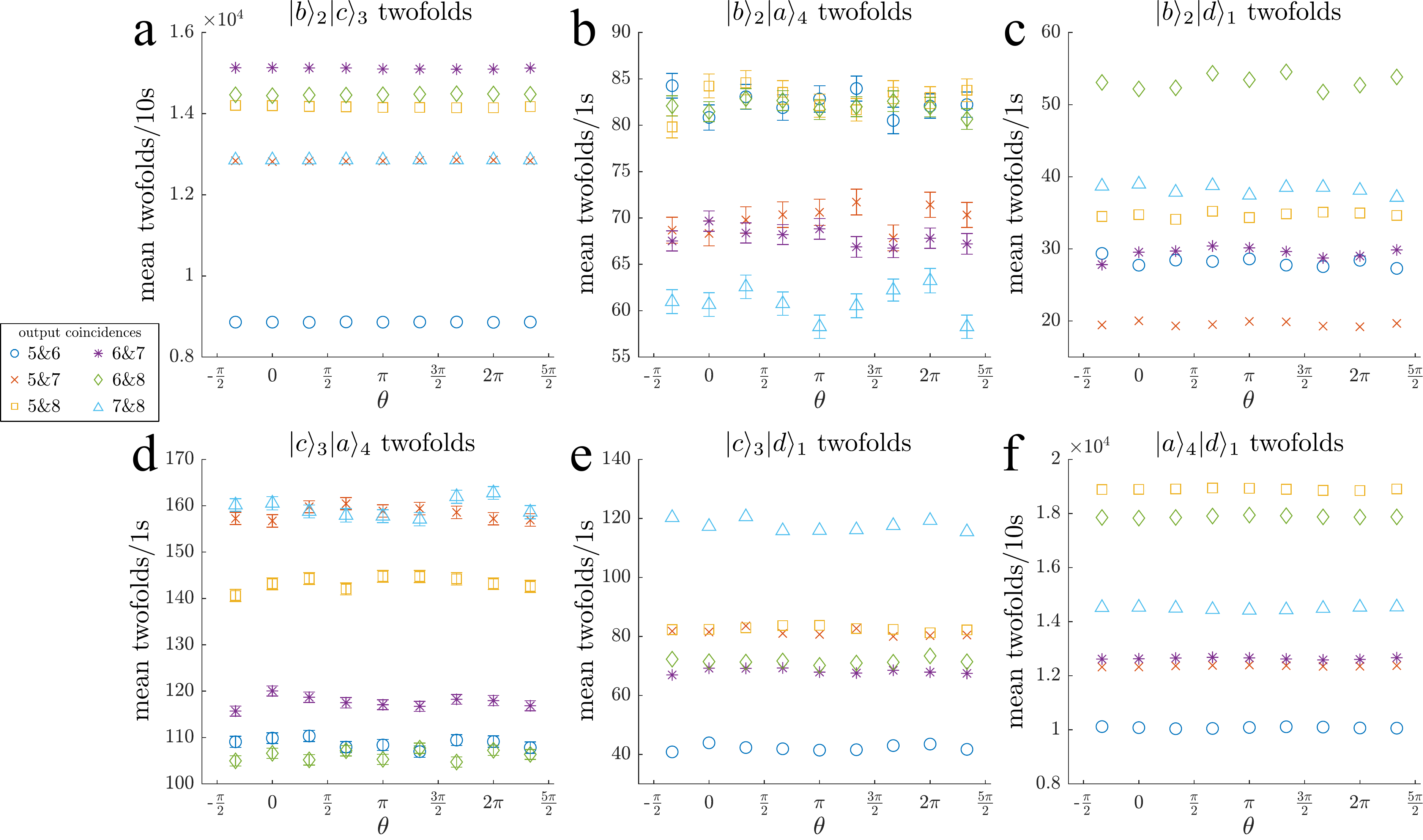}
	\caption{\label{fig:appTwofoldsCounts} Output twofold coincidences for each pair of quitter inputs recorded by blocking other arms and correcting for double emissions from a single source where appropriate. Error bars  smaller than markers are omitted. High counts correspond to twofolds from when a single source is used, and lower counts are twofolds between distinct sources. See main text for details.}		
\end{figure}

Next we average these normalised signals over all sweeps and also calculate the standard error on the mean value for each sampled $\theta$. Finally we multiply by the mean number of twofold coincidences recorded for $\theta=-\pi/3$ to allow comparison of count rates whilst not affecting the visibility of any variations. This allows direct association of variations in normalised twofold counts with changes in the relevant pairwise distinguishabilities, something that would otherwise be difficult to back out from signals arising due to the interference of independent thermal states, i.e. without this background-subtraction. Results are shown in Fig.~\ref{fig:appTwofoldsCounts}.

If our polarisation preparation is slightly imperfect then we expect any variations in the output twofolds to involve the rotating polarisation state of $\ket{d}$. In Fig.~\ref{fig:appTwofoldsCounts}f, the $\ket{a}_{4}\ket{d}_{1}$ twofolds correspond to counts when a single source ($\mathcal{B}$ in Fig.~\ref{fig:appSourceInterference}) is injected into the quitter. The average variation of the twofolds is under $0.2\%$. The largest variation of $0.8\%$ on outputs $7\&8$ has a $+\cos\theta$ shape that we attribute to a very small change in $r_{ad}^{2}$ as $\theta$ is varied. This could arise from a waveplate being $\sim0.4^{\circ}$ off the intended angle, meaning that the $H$ polarisation of $\ket{a}$ is not exactly perpendicular to the equatorial plane defined by $\ket{d}$ in the Bloch sphere. This small variation in $r_{ad}^{2}$ could cause a maximum of around $1\%$ variation in the associated double emission fourfolds for that source, and has a negligible impact on the background-subtracted fourfold signal visibility.

The other interference signals involving $\ket{d}$ are shown in Fig.~\ref{fig:appTwofoldsCounts}c,e, and these correspond to two-photon interference from distinct sources (see equation~\ref{eqn:appxTwoSourceNorm}). Those in Fig.~\ref{fig:appTwofoldsCounts}c correspond to interfering $\ket{b}$ and $\ket{d}$. The coincidences at pairs of outputs $k,l=(5,7),(5,8),(6,7),(6,8)$ are all predicted to have a dependence on $(1-r_{bd}^{2})$, analogous to usual HOM interference (this can be seen from Fig.~\ref{fig:appFoldedQuitter}a). Averaging the normalised signals for these output combinations suggests a visibility, and so maximum variation in $r_{bd}^{2}$, of around $2\%$. Using equation~\ref{eqn:bdOverlap} we then infer $\abs{\braket{t_{2}}{t_{3}}}\approx 0.15$. From equation~\ref{eqn:contributionRatio} we know that the ratio of four- to threefold contributions is $R^{(4/3)}\sim-1/\abs{\braket{t_{2}}{t_{3}}}$ and of four- to twofold contributions is $R^{(4/2)}\sim1/\abs{\braket{t_{2}}{t_{3}}}^{2}$. Therefore the four-photon term still by far contributes the majority of any $-\cos\theta$ variation. The variations in the other output pairs are about $1\%$, consistent with this value of temporal overlap magnitude. The twofolds in Fig.~\ref{fig:appTwofoldsCounts}e are all constant within error. Any small variations would again have a very small impact on the relative strengths of the four- and twofold contributions in $-\cos\theta$ to the background-subtracted $P_{1111}$ signal. The remaining signals in Fig.~\ref{fig:appTwofoldsCounts}a,b,d are all constant with $\varphi_{abcd}$. We conclude that any contributions to the four-photon coincidence probability from variations in pairwise distinguishabilities are very small.

\subsubsection{Three source arms open}
\label{appxsec:threeArms}
\begin{figure}[ht]
	\centering
	\includegraphics[width=0.9\textwidth]{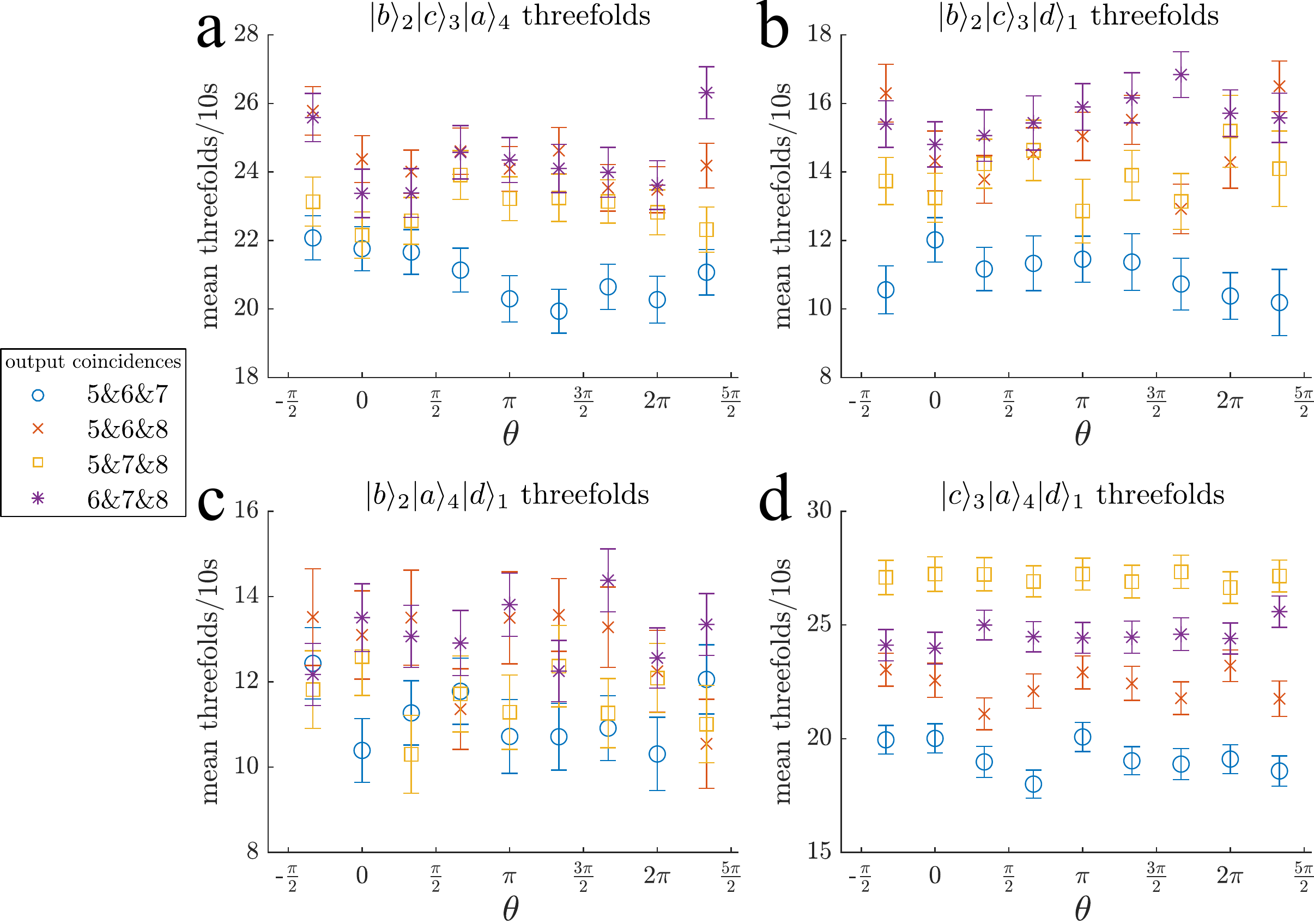}
	\caption{\label{fig:appThreefoldsCounts}Output threefold coincidences for each set of three quitter inputs, correcting for counts arising from a single source firing twice. Error bars are from repeated sweeps. See main text for details.}
\end{figure}

Once again we want to remove the effect of singles variations on threefold output coincidences recorded when three of the shutters are open. This configuration is equivalent to interfering a two-mode squeezed vacuum state from one source with a thermal state from the other unheralded source. The number of output threefolds recorded when opening inputs $i,j,k$ and monitoring outputs $l,m,n$ is
\begin{equation}
C^{i,j,k}_{l,m,n}(r_{ij}^{2},r_{ik}^{2},r_{jk}^{2},\theta)=R'_{\tau}\times\eta_{i}\eta_{j}\eta_{k}\eta_{l}\eta_{m}\eta_{n}\times P^{i,j,k}_{l,m,n}(r_{ij}^{2},r_{ik}^{2},r_{jk}^{2},\theta)+C_{3bg}.
\end{equation}
The three-photon scattering probability $P^{i,j,k}_{l,m,n}$ depends on the pairwise distinguishabilities, polarisation angle $\theta$ (through which any triad phase $\varphi_{ijk}$ is included) but not on the quitter phase $\chi$. The term denoted $C_{3bg}$ contains all output threefolds arising from cases where a single source (for example $\mathcal{A}$ on inputs $2,3$) fires twice and the other source does not fire. These are measured separately and can be subtracted to yield
\begin{equation}
C^{i,j,k}_{l,m,n,(bgsub)}(r_{ij}^{2},r_{ik}^{2},r_{jk}^{2},\theta)=R'_{\tau}\times\eta_{i}\eta_{j}\eta_{k}\eta_{l}\eta_{m}\eta_{n}\times P^{i,j,k}_{l,m,n}(r_{ij}^{2},r_{ik}^{2},r_{jk}^{2},\theta).
\end{equation}
Now we divide these background-subtracted threefolds by the relevant separately measured singles so
\begin{equation}
\begin{aligned}
\tilde{C}^{i,j,k}_{l,m,n,(bgsub)}(r_{ij}^{2},r_{ik}^{2},r_{jk}^{2},\theta)&=\frac{C^{i,j,k}_{l,m,n,(bgsub)}(r_{ij}^{2},r_{ik}^{2},r_{jk}^{2},\theta)}{C^{i}_{l}(\theta)\times C^{j}_{m}(\theta)\times C^{k}_{n}(\theta)}\\
&=\frac{R'_{\tau}}{R^{3}_{\tau}\times P^{i}_{l}(\theta)P^{j}_{m}(\theta)P^{k}_{n}(\theta)}\times P^{i,j,k}_{l,m,n,(bgsub)}(r_{ij}^{2},r_{ik}^{2},r_{jk}^{2},\theta)
\end{aligned}
\end{equation}
The effect of varying losses has been removed. As in the previous section, we do not infer three-photon scattering probabilities because of the high sensitivity to single-particle scattering probabilities. Instead we take the same approach as earlier: we divide each sweep's normalised signal by its mean over $\theta$ (only one of the single-particle scattering probabilities in the denominator could depend on $\theta$ if the input port is 1, so again variations from this term, when averaged over sampled $\theta$, are negligible). We then average across all sweeps and determine the standard error at each sampled $\theta$ value, then multiply by the mean counts at $\theta=-\pi/3$ to allow comparison of statistics without affecting visibilities. Results are shown in Fig.~\ref{fig:appThreefoldsCounts}.

Again only signals involving $\ket{d}$ are expected to vary with $\theta$. Given the possible variations of pairwise distinguishabilities inferred in the previous section, threefold signals involving state $\ket{d}$ are predicted to vary by at most around $2.5\%$. Due to low count rates for threefolds, error bars are quite large but all these signals are consistent with either being constant or with  this predicted modulation level. We therefore confirm that any three-photon interference contributions affecting the fourfold coincidence probability are much smaller than the four-photon terms.

\subsection{Four-photon coincidence signals}
\label{appxsec:singlesNormalisation}
\subsubsection{Background-subtracted fourfolds}
We can write the number of background-subtracted fourfolds obtained by subtracting the double-emission background counts from the total fourfold signal (see Supplementary~\ref{app:subtractingStatistics}) as
\begin{equation}
C^{1,2,3,4}_{5,6,7,8,(bgsub)}(\{r_{ij}^{2}\},\theta,\chi)=R'_{\tau}\times\eta_{1}\eta_{2}\eta_{3}\eta_{4}\eta_{5}\eta_{6}\eta_{7}\eta_{8}\times P^{\mathcal{AB}}_{1111}(\{r_{ij}^{2}\},\theta,\chi),
\end{equation}
where we have ignored six-photon terms and $P^{\mathcal{AB}}_{1111}(\{r_{ij}^{2}\},\theta,\chi)$ is the coincidence probability from equations~\ref{eqn:P1111} and ~\ref{eqn:appP1111}, depending on pairwise distinguishabilities $\{r_{ij}^{2}\}$, polarisation angle $\theta$ and the quitter phase $\chi$. Now dividing by the appropriate singles counts gives
\begin{equation}
\tilde{C}^{1,2,3,4}_{5,6,7,8,(bgsub)}(\{r_{ij}^{2}\},\theta,\chi)=\frac{C^{1,2,3,4}_{5,6,7,8}(\{r_{ij}^{2}\},\theta,\chi)}{C^{1}_{5}(\theta)C^{2}_{6}C^{3}_{7}C^{4}_{8}}=\frac{R'_{\tau}}{R^4_{\tau}\times P^{1}_{5}(\theta)P^{2}_{6}P^{3}_{7}P^{4}_{8}}\times P^{\mathcal{AB}}_{1111}(\{r_{ij}^{2}\},\theta,\chi).
\end{equation}
This normalises out the effect of varying coupling for each sweep of $\theta$. We normalise each sweep to its mean and average over all sweeps, before multiplying by the average counts at $\theta=-\pi/3$. Results when using all sampled quitter phase values $\chi$ are shown in Fig.~\ref{fig:fig4}f of the main text, along with discussion.

\subsubsection{Double-emission background fourfolds}
We perform a similar normalisation to singles counts for the separately measured double-emission background fourfolds. For example for those arising from source $\mathcal{A}$ we have
\begin{equation}
C^{2,2,3,3}_{5,6,7,8}(r_{bc}^{2},\chi)=R'_{\tau}\times\eta_{2}^{2}\eta_{3}^{2}\eta_{5}\eta_{6}\eta_{7}\eta_{8}\times P^{\mathcal{AA}}_{1111}(r_{bc}^{2},\chi).
\end{equation}
This scattering probability is given in equation~\ref{eqn:doubleEmissionCoincidences}. Next we normalise to products of singles as
\begin{equation}
\tilde{C}^{2,2,3,3}_{5,6,7,8}(r_{bc}^{2},\chi)=\frac{C^{2,2,3,3}_{5,6,7,8}(r_{bc}^{2},\chi)}{C^{2}_{5}\times C^{2}_{6}\times C^{3}_{7}\times C^{3}_{8}}=\frac{R'_{\tau}}{R^{4}_{\tau}\times P^{2}_{5}P^{2}_{6}P^{3}_{7}P^{4}_{8}}\times P^{\mathcal{AA}}_{1111}(r_{bc}^{2},\chi).
\end{equation}
This eliminates the effect of slow variations in coupling on the background fourfolds for each sweep of $\theta$. We perform this normalisation for source $\mathcal{B}$ too, and then do the same procedure as described for the background-subtracted signal above to find the normalised mean background fourfolds arising from individual sources firing twice. The resulting signals are shown in Fig.~\ref{fig:fig4}e of the main text, along with discussion.

\subsubsection{Total fourfolds}
The total fourfold signal measured when all source arms are open was presented in equations~\ref{eqn:Ptot} and~\ref{eqn:P1111tot2}, and is given by the sum of the signals from each source firing once, or one of the sources firing twice. The technique of normalising to separately measured singles counts therefore does not directly apply to this signal. Instead of summing the processed counts $\tilde{C}^{1,2,3,4}_{5,6,7,8,(bgsub)},\tilde{C}^{2,2,3,3}_{5,6,7,8}$ and $\tilde{C}^{1,1,4,4}_{5,6,7,8}$, we instead plot the mean of the raw measured total counts in Fig.~\ref{fig:fig4}d of the main text. This is almost identical to the signal that would result from summing the counts in Fig.~\ref{fig:fig4}e,f.

\subsection{Data for different postselected quitter phase ranges}
\label{app:backgroundFourfolds}
From equation~\ref{eqn:appP1111} and Table~\ref{table:exchangeContributions} we can see that magnitude of the four-photon exchange term in $\cos\varphi_{abcd}$ is controlled by the quitter phase through $(\cos 2\chi-2)$. In order to verify that it is this term that causes variations in the total and background-subtracted fourfold signals (Fig.~\ref{fig:fig4}d,f), we can investigate the dependence of their visibilities on $\chi$. The data presented in Fig.~\ref{fig:fig4} of the main text correspond to $P_{1111}$ sampled over known $\chi$ values determined using the method in Supplementary~\ref{app:measuringQuitterPhase}. We can instead postselect on different ranges of $\chi$ to verify the dependence of $P_{1111}$ visibility shown in Fig.~\ref{fig:appExchangeAndVisWithPhase}b.

\subsubsection{High visibility fourfold signal}
\label{appx:highVisSignal}
Consider postselecting for quitter phases in the range $\arccos\left(1/3\right)=1.23\leq\chi\leq\pi-\arccos\left(1/3\right)=1.91$ (corresponding to the central third of variations about $\chi=\pi/2$ in Fig.~\ref{fig:lockingDemo}). This means $(\cos 2\chi-2)\approx -3$, larger in magnitude than when averaging over all sampled values, so the fourfold signal visibility should increase. Including the effects of imbalanced input losses, higher-order emissions, and slight spectral distinguishability, the total and background-subtracted fourfold signal visibilities are predicted to be $8.5\%$ and $22.5\%$ respectively. Signals normalised using the procedure described in the previous subsections of this section of the Supplementary are shown in Fig.~\ref{fig:goodPhaseDataPlot}d,e, along with a selection of output singles, twofolds, threefolds, and the double emission backgrounds.

The variations in the singles, twofolds and threefolds in Fig.~\ref{fig:goodPhaseDataPlot}a,b,c are small enough that two- and three-photon contributions to the fourfold signal resulting from changes in pairwise distinguishability or triad phases are very small. The total fourfolds in Fig.~\ref{fig:goodPhaseDataPlot}d exhibit $-\cos\varphi_{abcd}$ variation with visibility $6.9\pm1.8\%$, consistent with the predicted $8.5\%$. Subtracting the flat backgrounds from double emissions shown in Fig.~\ref{fig:goodPhaseDataPlot}e yields the signal in Fig.~\ref{fig:goodPhaseDataPlot}f with visibility $23.6\pm7.0\%$, consistent with the predicted $22.5\%$. 

\begin{figure}[ht]
	\centering
	\includegraphics[width=0.8\textwidth]{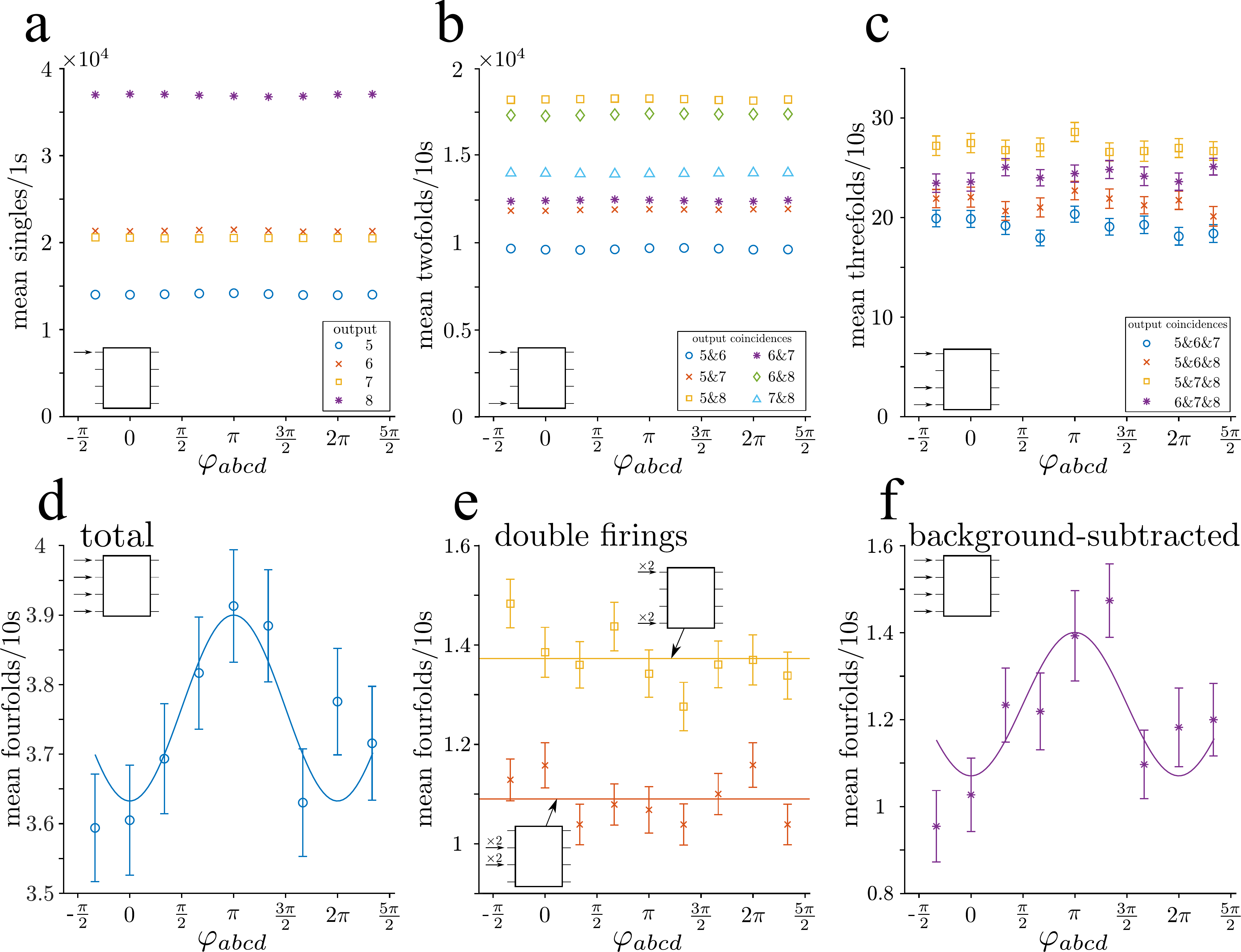}
	\caption{\label{fig:goodPhaseDataPlot} \textbf{a} Singles counts at the quitter outputs when injecting $\ket{d}_{1}$. \textbf{b} Output twofolds when injecting $\ket{d}_{1}\ket{a}_{4}$. \textbf{c} Output threefolds when injecting $\ket{d}_{1}\ket{c}_{3}\ket{a}_{4}$. \textbf{d} Output fourfold coincidences when injecting all source emissions into the quitter, corresponding to the total signal described in equation~\ref{eqn:P1111tot2}. The fitted cosine has a visibility of $6.9\pm1.8\%$ and the numbers of counts per point are between 2,477 and 2,275. \textbf{e} Output fourfolds arising from double emissions from the same source, measured by blocking each source in turn to inject $\ket{d}_{1}\ket{a}_{4}$ and $\ket{b}_{2}\ket{c}_{3}$. The mean numbers of counts per point are respectively 667 and 813. \textbf{f} Performing the background subtraction described in Supplementary~\ref{app:subtractingStatistics} yields the measured $P_{1111}$ signal. The fitted cosine has a visibility of $23.6\pm7.0\%$ and the number of counts per point is between 1,104 and 756.}
\end{figure}
\subsubsection{Low visibility fourfold signal}
\begin{figure}[ht]
	\centering
	\includegraphics[width=0.8\textwidth]{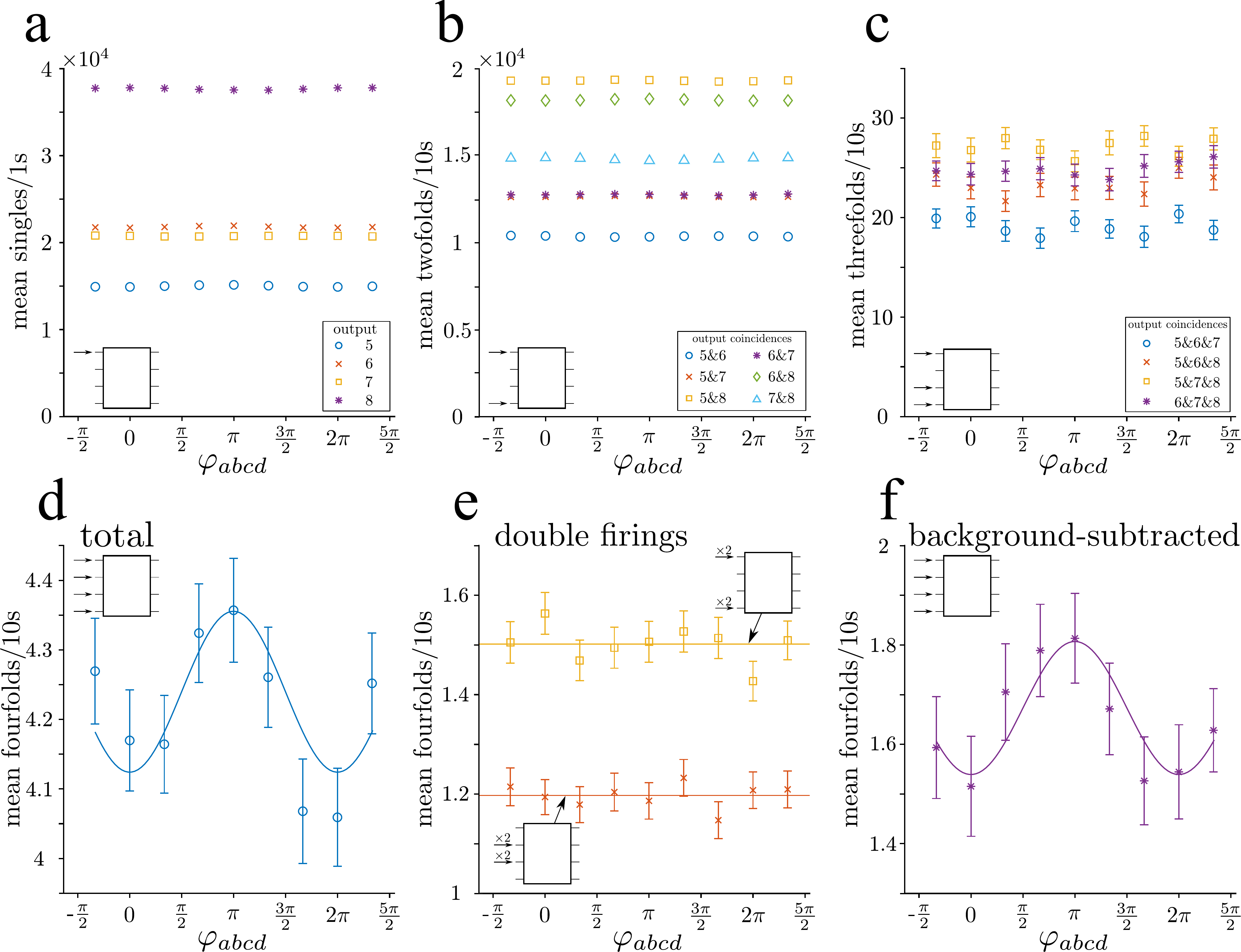}
	\caption{\label{fig:badPhaseDataPlot} \textbf{a} Singles counts at the quitter outputs when injecting $\ket{d}_{1}$. \textbf{b} Output twofolds when injecting $\ket{d}_{1}\ket{a}_{4}$. \textbf{c} Output threefolds when injecting $\ket{d}_{1}\ket{c}_{3}\ket{a}_{4}$. \textbf{d} Output fourfold coincidences when injecting all source emissions into the quitter, corresponding to the total signal described in equation~\ref{eqn:P1111tot2}. The fitted cosine has a visibility of $5.3\pm1.6\%$ and the numbers of counts per point are between 3,978 and 3,706. \textbf{e} Output fourfolds arising from double emissions from the same source, measured by blocking each source in turn to inject $\ket{d}_{1}\ket{a}_{4}$ and $\ket{b}_{2}\ket{c}_{3}$. The mean numbers of counts per point are respectively 1,072 and 1,348. \textbf{f} Performing the background subtraction described in Supplementary~\ref{app:subtractingStatistics} yields the measured $P_{1111}$ signal. The fitted cosine has a visibility of $14.8\pm3.0\%$ and the number of counts per point is between 1,630 and 1,381.}
\end{figure}

If we instead postselect on the remaining $\chi$ range corresponding to the top and bottom thirds of the variations in coincidences shown in Fig.~\ref{fig:lockingDemo} then $(\cos2\chi-2)\approx -1.75$ so the visibility should decrease. Including the effects of imbalanced input losses, higher-order emissions, and slight spectral distinguishability, the total and background-subtracted fourfold signal visibilities are predicted to be $4.5\%$ and $12.0\%$ respectively. The normalised measured signals are shown in Fig.~\ref{fig:badPhaseDataPlot}d,e, along with a selection of output singles, twofolds, threefolds, and the double emission backgrounds.

Again variations in the singles, twofolds and threefolds in Fig.~\ref{fig:goodPhaseDataPlot}a,b,c are small, and the double emission backgrounds in Fig.~\ref{fig:badPhaseDataPlot}e are constant with $\varphi_{abcd}$. Since the variations in fourfold signals shown in Fig.~\ref{fig:badPhaseDataPlot}d,f are so small, the fits are relatively poor yielding $5.3\pm1.6\%$ and $14.8\pm3.0\%$ for total and background-subtracted respectively. These are consistent with the predicted values and the variations are smaller than those when postselecting for the other $\chi$ range in the previous section. This confirms the $\chi$ dependence of the observed $\varphi_{abcd}$-dependent variations and hence the four-photon term from distinguishable state interference is what dominates the signals in Fig.~\ref{fig:fig4}d,f. 
\end{document}